\documentclass[11pt]{article}

\usepackage[margin=1in]{geometry}
\usepackage{amssymb}
\usepackage{booktabs}
\usepackage{tabularx}
\usepackage{multirow}
\usepackage{enumitem}
\usepackage{xcolor}
\usepackage{listings}
\usepackage{soul}
\usepackage[normalem]{ulem}
\usepackage{hyperref}
\usepackage{natbib}
\usepackage[section]{placeins}
\usepackage{graphicx}

\sethlcolor{yellow}

\begin{document}

\title{Inside the Scaffold: A Source-Code Taxonomy of Coding Agent Architectures}

\author{Benjamin Rombaut}
\date{}

\maketitle

\begin{abstract}
LLM-based coding agents can localize bugs, generate patches, and run tests
with diminishing human oversight, yet the scaffolding code that surrounds the
language model (the control loop, tool definitions, state management, and
context strategy) remains poorly understood. Existing surveys classify agents
by abstract capabilities (tool use, planning, reflection) that cannot
distinguish between architecturally distinct systems, and trajectory studies
observe what agents do without examining the scaffold code that determines
why. This paper presents a source-code-level architectural taxonomy derived
from analysis of 13 open-source coding agent scaffolds at pinned commit
hashes. Each agent is characterized across 12 dimensions organized into three
layers: control architecture, tool and environment interface, and resource
management. The analysis reveals that scaffold architectures resist discrete
classification: control strategies range from fixed pipelines to Monte Carlo
Tree Search, tool counts range from 0 to 37, and context compaction spans
seven distinct strategies. Five loop primitives (ReAct,
generate-test-repair, plan-execute, multi-attempt retry, tree search)
function as composable building blocks that agents layer in different
combinations; 11 of 13 agents compose multiple primitives rather than relying
on a single control structure. Dimensions converge where external constraints
dominate (tool capability categories, edit formats, execution isolation) and
diverge where open design questions remain (context compaction, state
management, multi-model routing). All taxonomic claims are grounded in file
paths and line numbers, providing a reusable reference for researchers
studying agent behavior and practitioners designing new scaffolds.
\end{abstract}

\section{Introduction}
\label{sec:intro}

Large language models have transformed software engineering practice. Tools such as
\textsf{Aider}~\citep{gauthier2024aider}, \textsf{SWE-agent}~\citep{yang2024sweagent},
and \textsf{OpenHands}~\citep{wang2024openhands} can navigate unfamiliar repositories,
localize bugs, generate patches, and run test suites with diminishing human oversight.
The most capable of these systems resolve over half of real-world GitHub issues on
the SWE-bench Verified benchmark~\citep{jimenez2024swebench}, and seven of the agents
analyzed in this study have each accumulated over 15,000 GitHub
stars~\citep{borges2018stars} (Table~\ref{tab:agents}), suggesting adoption by
substantial developer communities. As these agents move from
research prototypes to production tooling, the scaffolding code that surrounds the
language model (the control loop, tool definitions, state management, and context
strategy) increasingly determines how the agent behaves, what mistakes it makes, and
where it spends its token budget~\citep{ref3_opendev}.

Despite this practical significance, the internal architectures of coding agent scaffolds
remain poorly understood in the research literature. Existing surveys of LLM-based agents
operate at the conceptual level, organizing systems by abstract capabilities (tool use,
memory, planning, reflection) rather than by the implementation strategies that distinguish
one production system from another~\citep{masterman2024landscape, nowaczyk2025agentic}.
A coding agent that uses Monte Carlo Tree Search to explore candidate patches and one that
uses a simple while loop with test-driven retries both qualify as ``tool-using, planning,
reflective'' agents under these taxonomies, yet their scaffold architectures differ in
fundamental ways that affect cost, reliability, and failure modes. Empirical work on coding
agents has begun to characterize their runtime behavior through trajectory
analysis~\citep{majgaonkar2025trajectories, ceka2025traceability}, revealing that
successful agents localize bugs faster, test earlier, and produce shorter action sequences
than failing ones. However, these studies treat agents as black boxes: they observe what
agents do, but do not examine the scaffold code that determines why. Researchers have
called for architecture-aware evaluation metrics that link internal components to
observable outcomes~\citep{ref4_archaware}, and detailed architectural descriptions exist
for individual systems~\citep{ref3_opendev}, but, to the best of the current literature
search, no study has systematically compared the scaffolding architectures of production
coding agents at the source-code level.

This gap matters for three reasons. First, without a shared vocabulary for scaffold design,
researchers studying agent behavior cannot attribute observed differences to specific
architectural choices; the confound between scaffold design and model capability goes
unacknowledged. Prior trajectory studies used different LLMs for different
agents~\citep{majgaonkar2025trajectories}, making it impossible to isolate whether a
behavioral difference stems from the scaffold or the model. Second, practitioners building
new coding agents lack a systematic map of the design space. Architectural decisions
(whether to use a persistent event store or a flat message list, whether to give the LLM
ten specialized tools or a single shell command, whether to compact context via
summarization or truncation) are currently made by reading individual codebases or blog
posts, with no comparative reference (Section~\ref{sec:related}). Third, the rapid pace of development means the
design space is expanding faster than it is being documented: the 13 agents analyzed in
this study span release dates from June 2023 to March 2025 and represent
distinct architectural strategies (fixed pipelines, sequential ReAct loops, phased
workflows, depth-first tree search, and full MCTS), yet no prior work has mapped their
relationships.

This paper presents a source-code-level architectural taxonomy of 13 open-source coding
agent scaffolds. Each agent's implementation was analyzed at a pinned commit hash across
12 dimensions organized into three layers: control architecture (how the agent decides
what to do next), tool and environment interface (how the agent interacts with code and
execution environments), and resource management (how the agent manages context, state,
and models) (Section~\ref{sec:methodology:dimensions}).
The analysis follows a qualitative case-study methodology, with every
taxonomic claim grounded in file paths and line numbers from cloned repositories.

The taxonomy reveals several findings that challenge common assumptions about coding agent
design. Rather than falling into discrete architectural categories, agents occupy positions
along continuous spectra: control strategies range from fixed pipelines with no feedback
loop (\textsf{Agentless}~\citep{xia2024agentless}) to full Monte Carlo Tree Search with
reward backpropagation (\textsf{Moatless Tools}~\citep{moatlesstools}), with 7 of 13 agents using a
sequential ReAct loop as their primary control structure. Loop primitives (ReAct,
generate-test-repair, plan-execute, multi-attempt retry) function as composable building
blocks that agents layer in different combinations, not as mutually exclusive types. Tool
sets vary from zero LLM-callable tools (\textsf{Aider}, where the user drives all
navigation) to 37 action classes (\textsf{Moatless Tools}), yet the underlying capability
categories converge: reading, searching, editing, and executing code appear in every
agent that grants the LLM autonomy. Context compaction, state management, and
multi-model routing each exhibit similar spectra, with design choices at one end
prioritizing simplicity and at the other prioritizing robustness or search breadth.
Five cross-cutting themes, including the tradeoff between sampling and iteration, the
emergence of sub-agent delegation as a first-class tool, and the architectural
implications of IDE coupling, span multiple dimensions and capture patterns that do not
reduce to a single axis.

The contributions of this paper are as follows:

\begin{enumerate}[leftmargin=*]
  \item A taxonomy of coding agent scaffold architectures derived from source-code
    analysis of 13 open-source agents, organized into three layers and 12 dimensions.
    This is, to the best of the current literature search, the first comparative
    architectural study of coding agents at the implementation level.
  \item The empirical finding that scaffold architectures are better characterized as
    compositions of loop primitives along continuous spectra than as instances of
    discrete architectural types.
  \item A detailed evidence base of architectural observations pinned to specific commits,
    providing a reusable reference for researchers studying agent behavior and
    practitioners designing new scaffolds.
\end{enumerate}

The remainder of this paper is structured as follows.
Section~\ref{sec:related} surveys related work on agent architectures, trajectory
analysis, and coding agent evaluation.
Section~\ref{sec:methodology} describes the agent selection criteria, analysis dimensions,
and coding procedure.
Section~\ref{sec:results} presents the taxonomy across its three layers, with evidence
tables and cross-cutting themes.
Section~\ref{sec:discussion} interprets the findings and draws out implications for
agent design and evaluation.
Section~\ref{sec:threats} addresses threats to validity.
Section~\ref{sec:conclusion} concludes with a summary and directions for future work.

\section{Related Work}
\label{sec:related}

This study sits at the intersection of several research threads. The following
subsections address each in turn: conceptual taxonomies of LLM agents
(Section~\ref{sec:related:surveys}), empirical studies of coding agent behavior
(Section~\ref{sec:related:trajectories}), architectural descriptions of
individual systems (Section~\ref{sec:related:systems}), and benchmark-driven
evaluation (Section~\ref{sec:related:benchmarks}). Each thread contributes a
different perspective on coding agents; none examines their scaffold
architectures comparatively at the source-code level.

\subsection{LLM Agent Architecture Surveys}
\label{sec:related:surveys}

A growing body of survey work organizes LLM-based agents by abstract
capabilities. \citet{masterman2024landscape} propose a
five-component reference model (reasoning, planning, tool use, memory,
reflection) and classify interaction patterns across single-agent and
multi-agent designs. \citet{nowaczyk2025agentic} offers a
capability-based taxonomy (tool-using, memory-augmented, planning,
multi-agent, embodied) grounded in classical agent theory, arguing that
reliability is an architectural property rather than a function of model
quality alone. This claim is sympathetic to the premise of the present study;
the source-code analysis presented here provides the implementation-level
evidence that Nowaczyk et~al.'s conceptual argument implies but does not
supply, grounding the relationship between architecture and reliability in
specific scaffold design choices rather than abstract capability categories.
Broader surveys of LLM applications in software engineering
cover coding agents as one category among many~\citep{zhang2026survey}. These
taxonomies provide useful conceptual vocabulary, but they operate at a level of
abstraction that cannot distinguish between production coding agents. Every
agent in the present corpus qualifies as ``tool-using, memory-augmented,
planning'' under these schemes, yet their scaffold architectures differ in ways
that affect cost, reliability, and failure modes. A coding agent that uses
Monte Carlo Tree Search to explore candidate patches and one that uses a simple
while loop with test-driven retries are indistinguishable under capability-based
classification, despite fundamental differences in control flow, state
management, and resource consumption. A related strand of work studies how LLMs
learn to use tools at the model level (tool-calling training, function-calling
fine-tuning)~\citep{schick2023toolformer, patil2024gorilla}; this addresses a complementary question to ours, since the
present study examines how scaffold code \emph{exposes and orchestrates} tools,
not how models learn to invoke them.

The prompting paradigms that underlie many of these conceptual categories have
been formalized independently.
\citet{yao2023react} introduce the \emph{ReAct} paradigm, in which
LLMs generate interleaved reasoning traces and task-specific actions in a
thought-action-observation loop; seven of the 13~agents in this study
use a sequential ReAct loop as their primary control structure
(Table~\ref{tab:control-loop}), and several others embed ReAct-like
iteration within phases of a larger workflow. \citet{shinn2023reflexion}
formalized \emph{Reflexion}, a verbal reinforcement learning framework where
agents reflect on failed attempts and store reflections in an episodic memory
buffer for subsequent retries; this pattern is foundational to the
generate-test-repair loop primitive observed across the corpus
(Section~\ref{sec:results:loop-spectrum}). However, both ReAct and Reflexion
describe \emph{algorithmic} paradigms, not scaffold architectures. The gap
between ``interleave thoughts and actions'' and a production implementation
with tool registration, context compaction, multi-model routing, and persistent
state is precisely what this study documents.

\subsection{Coding Agent Trajectory and Behavior Studies}
\label{sec:related:trajectories}

A complementary line of work analyzes what coding agents \emph{do} at runtime
by studying their execution trajectories. \citet{ceka2025traceability} collect full execution traces from five agents
running on SWE-bench Verified, normalize them into a unified action schema, and
extract a taxonomy of ``decision pathways'': recurring behavioral patterns such
as exploration-heavy, patch-first, and test-driven strategies. They find that
bug localization is the primary bottleneck and that early test generation
correlates strongly with success. \citet{majgaonkar2025trajectories} analyze trajectory logs from
\textsf{OpenHands}, \textsf{SWE-agent}, and
\textsf{Prometheus}~\citep{prometheus},
reporting that
failure trajectories are 12--82\% longer than successful ones and that
repository navigation dominates agent activity over patch writing.
\citet{bouzenia2025tar} study thought-action-result
trajectories from three agents (\textsf{RepairAgent}~\citep{bouzenia2025repairagent}, \textsf{AutoCodeRover},
\textsf{OpenHands}), finding that even rare thought-action misalignment
(0.5--4.8\% of steps) strongly correlates with failure and that failing
trajectories exhibit repetitive, non-adaptive action cycles.

These studies establish important empirical regularities, but they share two
limitations that the present work addresses. First, they treat agents as black
boxes: they observe \emph{what} agents do but cannot explain \emph{why}. A
finding such as ``agents with shorter trajectories succeed more often'' cannot
distinguish between a scaffold that enforces early termination, one that
implements efficient search heuristics, and one that simply has a low iteration
budget. Second, prior trajectory studies used different LLMs for different
agents~\citep{majgaonkar2025trajectories, bouzenia2025tar}, confounding
scaffold effects with model effects; for example, Majgaonkar et al.\ compare
Claude~3.5 Sonnet-based \textsf{OpenHands} trajectories against
DeepSeek-V3-based \textsf{Prometheus} trajectories, making it impossible to
isolate whether behavioral differences stem from the scaffold or the model.
The architectural taxonomy presented here
provides the missing explanatory layer: by examining the scaffold source code
that produces those trajectories, it becomes possible to attribute behavioral
differences to specific design choices rather than to opaque system-level
differences.

\citet{fan2025sweeffi} move from behavioral to resource-oriented
analysis, measuring token consumption, cost, and computation time for five
scaffolds across three LLMs. Their ``token snowball effect'' (linear input
token growth with API calls due to naive conversation history accumulation) and
``expensive failures'' (failing attempts consuming up to 4$\times$ the
resources of successful ones) are empirical phenomena that the present study
can explain architecturally: the token snowball maps to the context compaction
dimension (Section~\ref{sec:results:compaction}), and expensive failures relate to
error recovery and termination strategies. However, \textsf{SWE-Effi} does not
examine the scaffold code that produces these cost patterns; its only
architectural distinction is ``agentic'' versus ``procedural.''

\subsection{Individual System Descriptions}
\label{sec:related:systems}

Detailed architectural descriptions exist for several individual coding agents,
and collectively they demonstrate that scaffold design choices significantly
affect agent behavior. Several papers focus on how tool and environment
interfaces shape agent capabilities: \citet{yang2024sweagent}
introduce the agent-computer interface (ACI) concept for \textsf{SWE-agent},
showing that designing custom shell commands to structure repository
interaction is itself an architectural decision that affects downstream
performance; \citet{gauthier2024aider} describes repository mapping
via PageRank and model-specific edit format selection in \textsf{Aider};
and \citet{ref3_opendev} describes a four-layer terminal agent architecture
with dual-agent design, multi-model routing, lazy tool discovery, and adaptive
context compaction, concluding that ``tool reliability matters more than model
capability.''

Other system papers illuminate the spectrum of agent autonomy. At one end,
\citet{xia2024agentless} argue that agentic scaffolds are
unnecessary for many tasks, demonstrating that a fixed pipeline of
localization, repair, and re-ranking stages can match or exceed agentic
approaches on SWE-bench; this motivates the pipeline-to-agent spectrum in the
present taxonomy. At the other end, \citet{aggarwal2025dars}
present tree-structured search with LLM-based branch selection in
\textsf{DARS-Agent}, and \citet{arora2024masai} decompose the
agent into specialized sub-agents for distinct subtasks in \textsf{MASAI}.
Between these extremes, \citet{zhang2024autocoderover} describe
AST-based context retrieval with spectrum-based fault localization in
\textsf{AutoCodeRover}, and \citet{wang2024openhands} describe the
event-sourced architecture and agent delegation mechanisms underlying
\textsf{OpenHands}. Each of these papers provides architectural depth for a
single system, but the design space as a whole remains unmapped: practitioners
cannot compare control loop strategies, tool interface designs, or context
management approaches without reading a dozen or more codebases independently.
The present study extends this per-system depth across 13~agents, enabling the
comparative analysis that individual descriptions cannot provide.

A related but orthogonal line of work studies how developers configure coding
agents. \citet{ref12_configuring} analyze 2,926 GitHub repositories to
characterize configuration artifacts (instruction files, prompt templates,
skills definitions, structured configuration) across five commercial coding
tools, finding that configuration practices remain fragmented and that advanced
features such as skills and sub-agent configurations are rarely used. This work
is complementary to the present study: configuration artifacts control what
instructions the agent receives, while scaffold architecture determines how the
agent processes those instructions. A \texttt{CLAUDE.md} file that specifies
``always run tests before committing'' is a developer-facing configuration; the
scaffold's generate-test-repair loop that implements that instruction is an
architectural feature. The present study analyzes the latter.

\subsection{Coding Agent Evaluation and Benchmarks}
\label{sec:related:benchmarks}

SWE-bench~\citep{jimenez2024swebench} has become the de facto evaluation standard
for coding agents, driving rapid progress and motivating the design of many
agents in the present corpus. However, its limitations are increasingly well
documented: overly detailed issue descriptions inflate resolution
rates~\citep{garg2026saving}, single-language bias limits generalizability~\citep{jimenez2024swebench, xu2025swecompass}, and
confounded scaffold-model effects make it difficult to attribute performance to
architectural choices~\citep{fan2025sweeffi}. Newer benchmarks address subsets of these concerns:
SWE-bench Pro~\citep{deng2025swebenchpro} introduces harder, multi-file tasks
with contamination resistance, and
SWE-Compass~\citep{xu2025swecompass} extends evaluation to eight programming
languages and multiple task types. \citet{chen2026rethinking}
examine whether agent-generated tests improve resolution rates under a minimal
scaffold, finding that prompt interventions that add or remove testing change
outcomes by at most 2.6~percentage points; this suggests that scaffold-level
orchestration of testing (lint-test cycles, test-gated retries, MCTS reward
signals), rather than model-native test-writing behavior, may be the
architecturally relevant variable.

The present study deliberately does not benchmark agent performance, because
benchmark scores confound scaffold architecture with model capability, prompt
engineering, and incidental configuration choices (iteration limits, cost caps,
default model selection); isolating the scaffold's contribution requires the
kind of architectural analysis presented here, not additional benchmark runs.
The taxonomy does, however, enable future controlled experiments by
identifying the specific variables that would need to be held constant: for
example, comparing agents with identical tool sets but different loop
strategies, or identical loops but different compaction strategies, with the
model held constant (Section~\ref{sec:discussion:evaluation}).

\citet{ref4_archaware} explicitly call for
architecture-aware evaluation metrics that link internal agent components
(planner, memory, tool router) to observable outcomes, proposing a
component-to-metric mapping framework. However, their proposal is conceptual:
it has not been tested on real systems and depends on architectural
documentation that, prior to this study, did not exist for most coding agents.
The present taxonomy provides the architectural vocabulary that such evaluation
frameworks require.

\section{Methodology}
\label{sec:methodology}
This study performs a source-code-level architectural analysis of open-source coding agent
scaffolds, deriving taxonomic categories from observed implementation patterns rather than
from documentation claims or conceptual models.

\subsection{Agent Selection}
\label{sec:methodology:selection}

Candidate agents were identified through three channels: agents evaluated on SWE-bench and reported
in the leaderboard literature as of early 2026, agents with substantial adoption in developer
tooling (measured by GitHub stars as a proxy for community use~\citep{borges2018stars}), and agents cited in recent surveys
of LLM-based software engineering~\citep{zhang2026survey}. The search was not exhaustive
within any single channel; the pool reflects the agents the researcher encountered through
these sources rather than a complete enumeration of all agents meeting a fixed threshold.
An initial pool of 22 candidates was narrowed using three inclusion criteria (the full
candidate list and exclusion rationale for each are provided in
Appendix~\ref{sec:appendix:candidates}):

\begin{enumerate}[leftmargin=*]
  \item \textbf{Coding-specific.} The agent must be designed for software engineering tasks
    (code editing, bug fixing, repository navigation), not general-purpose task automation.
    This excluded general-purpose frameworks
    (Open Interpreter~\citep{openinterpreter},
    Deep Agents~\citep{deepagents}) and multi-agent
    orchestration platforms (MetaGPT~\citep{hong2024metagpt},
    CrewAI~\citep{crewai}), whose unit of analysis
    is agent coordination rather than individual agent architecture.
  \item \textbf{Open source with readable implementation.} The agent's scaffolding code
    (control loop, tool definitions, state management) must be available as readable source
    code in a public repository, pinned to a specific commit. This excluded Claude
    Code~\citep{claudecode}, which is distributed as a compiled binary with no published
    source repository, and
    MASAI~\citep{arora2024masai}\footnote{The MASAI repository contains only a README linking
    to the paper; no implementation code has been released.}. Proprietary agents whose
    scaffold code is not inspectable (GitHub Copilot Workspace, Cursor's AI backend,
    Windsurf) were also excluded on this basis.
  \item \textbf{Architecturally distinct.} Near-duplicate agents were removed to avoid redundant
    analysis (e.g., agents sharing the same core codebase or differing only in frontend integration).
\end{enumerate}

The 13 agents retained for analysis are listed in Table~\ref{tab:agents}. They span two natural
origin categories: \emph{CLI agents} built for interactive developer use, and \emph{SWE-bench
agents} built primarily for automated issue resolution. One additional agent,
\textsf{mini-swe-agent}~\citep{minisweagent},
is included as a deliberate minimal baseline: it was released by the
\textsf{SWE-agent}~\citep{yang2024sweagent} team as a reference implementation exposing the simplest
possible complete scaffold. The selection is not exhaustive but aims to cover the range of
architectural strategies present in the open-source coding agent ecosystem as of early 2026.

\begin{table}[htbp]
  \caption{Agents analyzed, ordered by GitHub stars as a proxy for community
  adoption~\citep{borges2018stars}.
  ``Category'' distinguishes interactive CLI tools from agents designed for automated SWE-bench
  evaluation. All agents were analyzed at pinned commit hashes (see Appendix~\ref{sec:appendix:commits}).}
  \label{tab:agents}
  \centering\footnotesize
  \begin{tabular}{lllrl}
    \toprule
    \textbf{Agent} & \textbf{Category} & \textbf{Language} & \textbf{Stars} & \textbf{Origin} \\
    \midrule
    \textsf{OpenCode}~\citep{opencode}
      & CLI & TypeScript & 135k & SST \\
    \textsf{Gemini CLI}~\citep{geminicli}
      & CLI & TypeScript & 100k & Google \\
    \textsf{Codex CLI}~\citep{codexcli}
      & CLI & Rust / TS & 72k & OpenAI \\
    \textsf{OpenHands}~\citep{wang2024openhands}
      & SWE-bench & Python & 70k & All Hands AI \\
    \textsf{Cline}~\citep{cline}
      & CLI & TypeScript & 60k & Independent \\
    \textsf{Aider}~\citep{gauthier2024aider}
      & CLI & Python & 43k & Independent \\
    \textsf{SWE-agent}~\citep{yang2024sweagent}
      & SWE-bench & Python & 19k & Princeton / CMU \\
    \textsf{mini-swe-agent}~\citep{minisweagent}
      & Baseline & Python & 4k & Princeton / CMU \\
    \textsf{AutoCodeRover}~\citep{zhang2024autocoderover}
      & SWE-bench & Python & 3k & NUS / SonarSource \\
    \textsf{Agentless}~\citep{xia2024agentless}
      & SWE-bench & Python & 2k & UIUC \\
    \textsf{Prometheus}~\citep{prometheus}
      & SWE-bench & Python & 1k & EuniAI \\
    \textsf{Moatless Tools}~\citep{moatlesstools}
      & SWE-bench & Python & 600 & Independent \\
    \textsf{DARS-Agent}~\citep{aggarwal2025dars}
      & SWE-bench & Python & 70 & Independent \\
    \bottomrule
  \end{tabular}
\end{table}

\subsection{Analysis Dimensions}
\label{sec:methodology:dimensions}

The analysis framework covers nine dimensions (which yield 12 taxonomy
dimensions in the results, as described below).
These dimensions were not fixed a priori;
they emerged through iterative open coding~\citep{strauss1998basics} of source code during a
pilot analysis of two architecturally contrasting agents: \textsf{Aider} (a simpler, interactive CLI
scaffold) and \textsf{OpenHands} (a complex, event-driven, containerized scaffold). These two were
chosen to maximize architectural diversity during piloting; the rationale and specific
refinements are described in Section~\ref{sec:methodology:procedure}.

The initial pilot began with six candidate dimensions drawn from the conceptual agent
architecture literature~\citep{masterman2024landscape}: control loop, tool interface, state
management, context retrieval, execution isolation, and multi-model routing. During the pilot,
three additional dimensions were added after the source code revealed architectural variation
not captured by the initial set: tool discovery strategy (prompted by differences in how \textsf{Aider}
and \textsf{OpenHands} register tools), context compaction (prompted by \textsf{Aider}'s explicit summarization
logic versus \textsf{OpenHands}' lack thereof), and persistent memory (prompted by \textsf{Aider}'s conventions
files). Stabilization consisted of applying the revised nine-dimension framework to both pilot
agents a second time to confirm that every dimension produced discriminating findings across
both agents and that no source code observations fell outside the framework without being
captured by the open-ended section. No dimensions were removed during this process. Several of
the resulting dimensions align with architectural concerns identified independently in prior
single-system analyses of coding agents~\citep{ref3_opendev} and conceptual agent architecture
surveys~\citep{masterman2024landscape}, providing external support for their relevance. Each
dimension captures a distinct architectural decision point in the scaffold.

\begin{enumerate}[leftmargin=*]

  \item \textbf{Control loop type.} The structure of the agent's main execution loop: how it
    sequences LLM calls, tool dispatches, and observation handling. A sub-property, \emph{loop
    driver}, records whether the loop is user-initiated or LLM-driven, a distinction added
    during piloting after observing that it determines several downstream architectural
    decisions (Section~\ref{sec:results:loop-driver}).

  \item \textbf{Tool set and tool interface design.} The full set of tools available to the model,
    including how tools are defined and communicated to the LLM (e.g., JSON schema via function
    calling, inline system-prompt descriptions, or custom text formats). This dimension also records
    the edit and patch format used when the agent proposes file modifications.

  \item \textbf{Tool discovery strategy.} Whether the full tool set is registered at startup
    (static) or whether tools are conditionally loaded, dynamically registered, or assembled at
    runtime based on task state (dynamic).

  \item \textbf{State management strategy.} How the agent accumulates and represents history across
    loop iterations: the data structure (flat message list, typed event log, tree), whether state is
    append-only or mutable, and what is stored beyond raw messages.

  \item \textbf{Context retrieval paradigm.} The mechanism by which the agent identifies relevant
    source code before or during task execution: LLM-directed tool calls (\textsf{grep},
    \textsf{find}, AST queries),
    pre-computed repository indexes, embedding-based semantic search, or static repository maps
    included in the initial prompt.

  \item \textbf{Execution isolation model.} Where the agent runs shell commands and code: on the
    host filesystem, in a Docker container, in a sandboxed subprocess, or in a remote cloud
    environment. This records the trust boundary between the agent and the host system.

  \item \textbf{Context compaction approach.} The strategy used when conversation history approaches
    the model's context limit: hard truncation, sliding window, LLM-generated summarization,
    selective dropping of tool results, or no mechanism present.

  \item \textbf{Multi-model routing.} Whether a single model handles all agent steps or whether
    different models are assigned to different roles (e.g., a large model for planning, a small model
    for localization), and how routing decisions are made.

  \item \textbf{Persistent memory.} Whether any information survives between sessions: project
    conventions, prior task outcomes, user preferences, or repository facts, and the storage
    mechanism used.

\end{enumerate}

\noindent The analysis template also includes an open-ended tenth section for observations that do
not fit any of the nine dimensions. This follows standard practice in qualitative coding, where
emergent categories are expected alongside predefined ones~\citep{saldana2021coding}: forcing every
finding into a predefined category risks suppressing novel patterns. In practice, this section
captured 47 cross-cutting findings (enumerated in the per-agent analysis
documents) that informed the taxonomy
structure presented in Section~\ref{sec:results}.

The nine analysis dimensions map to 12 taxonomy dimensions in the results.
During analysis, two sub-properties proved sufficiently discriminating to
warrant independent treatment: \emph{loop driver} (a sub-property of control
loop type) and \emph{edit and patch format} (a sub-property of tool set
design) each produced distinct spectra with their own evidence tables and
are presented as separate dimensions in Section~\ref{sec:results}. A third
dimension, \emph{control flow implementation} (the code-level mechanism
realizing the control loop: while loop, recursion, compiled graph, or
exception-based signaling), emerged during analysis as an axis of variation
orthogonal to loop topology and is likewise presented independently. This
expansion from analysis framework to taxonomy structure is a normal outcome
of qualitative coding: the framework guides data collection, but the final
categories reflect what the data reveals~\citep{saldana2021coding}.

\subsection{Tool Counting Methodology}
\label{sec:methodology:toolcounting}

Tool counts are reported using two complementary methods to avoid conflating interface granularity
with functional coverage.

\textbf{Registration count} tallies tools as the LLM sees them: each separately registered callable
(function calling schema entry, system-prompt-defined command, or equivalent) is counted once. This
measure is sensitive to how scaffold developers chose to partition functionality.

\textbf{Capability category count} groups tools by what they do: read, search, edit, execute,
validate, and repository state. These categories were derived inductively from the pilot analysis:
after cataloguing every tool across the first two agents, recurring functional roles were grouped
into categories, which were then validated against the remaining eleven agents. This measure is
comparable across agents regardless of granularity choices. An agent with a single \textsf{bash}
tool and an agent with separate \textsf{run\_command}, \textsf{run\_tests}, and
\textsf{run\_linter} tools both cover the execute category. Both counts are reported in the
results; the capability category count is used for cross-agent comparison.

\subsection{Analysis Procedure}
\label{sec:methodology:procedure}

Each agent was analyzed using a structured template derived from the nine dimensions
established in Section~\ref{sec:methodology:dimensions}, applied consistently across all
13 agents.
The template separates three levels of description, following the principle that empirical
claims should be traceable to primary data sources rather than inferred from secondary
documentation~\citep{runeson2009guidelines}:

\begin{itemize}[leftmargin=*]
  \item \textbf{Observation}: what the code does, described in terms of data flow and control flow
    with specific file and line references.
  \item \textbf{Classification}: how the observed behavior maps to a taxonomy dimension, with
    explicit justification for the mapping.
  \item \textbf{Evidence}: a file path and line number pinned to the specific commit hash analyzed.
\end{itemize}

\noindent For each agent, the template produces a structured document covering all nine dimensions
plus the open-ended section, with a full call-chain trace of the main control loop as the entry
point. The template was piloted on the same two agents used for dimension development
(Section~\ref{sec:methodology:dimensions}): \textsf{Aider} and \textsf{OpenHands}. This pilot served a distinct
purpose from the dimension stabilization; it tested whether the three-level observation format
produced consistent and sufficiently detailed analysis documents. The pilot also surfaced the
need for a dual tool counting methodology
(Section~\ref{sec:methodology:toolcounting}) after observing that raw tool counts were not
comparable across agents with different interface granularity. The complete analysis documents
for all 13 agents follow this template; individual sections of
each analysis are referenced throughout the results where they serve as primary evidence for
specific claims.

All analyses were pinned to specific git commit hashes, listed in
Appendix~\ref{sec:appendix:commits}, because several agents under study (\textsf{Cline}, \textsf{Aider},
\textsf{OpenHands}, \textsf{Gemini CLI}) were under active development during the analysis period and unpinned
analysis would yield unreproducible results. Where source code was ambiguous, the analysis
records uncertainty explicitly rather than assigning a confident classification. Dimensions
that genuinely do not apply to an agent (for example, persistent memory in agents with no
inter-session storage) are recorded as absent, not omitted. All analyses were conducted by
the author with substantial use of LLM-based coding assistants for code navigation,
call-chain tracing, and initial summarization of unfamiliar codebases. All
LLM-generated observations were verified against the source code before inclusion in the
analysis documents; the analysis documents themselves
record file paths and line numbers to enable independent verification. The single-author
design is a threat to construct validity; mitigations are discussed in
Section~\ref{sec:threats}.

\subsection{Scope and Limitations}
\label{sec:methodology:scope}

This study is purely taxonomic. No performance benchmarking was conducted, and no claims are made
about correlations between scaffold design and task success rates. The decision to exclude
performance analysis reflects two limitations in the available data: first, SWE-bench results
across agents are not directly comparable because different agents used different underlying models,
and model capability is a larger confounder than scaffold design; second, documented solution
leakage in SWE-bench issue descriptions~\citep{garg2026saving} makes raw pass rates an unreliable signal
regardless of these controls. The corpus is limited to open-source agents with readable source
code; threats arising from this restriction are discussed in Section~\ref{sec:threats}.

\section{Results}
\label{sec:results}

\begin{figure*}[!htbp]
\centering
\includegraphics[width=0.5\textwidth]{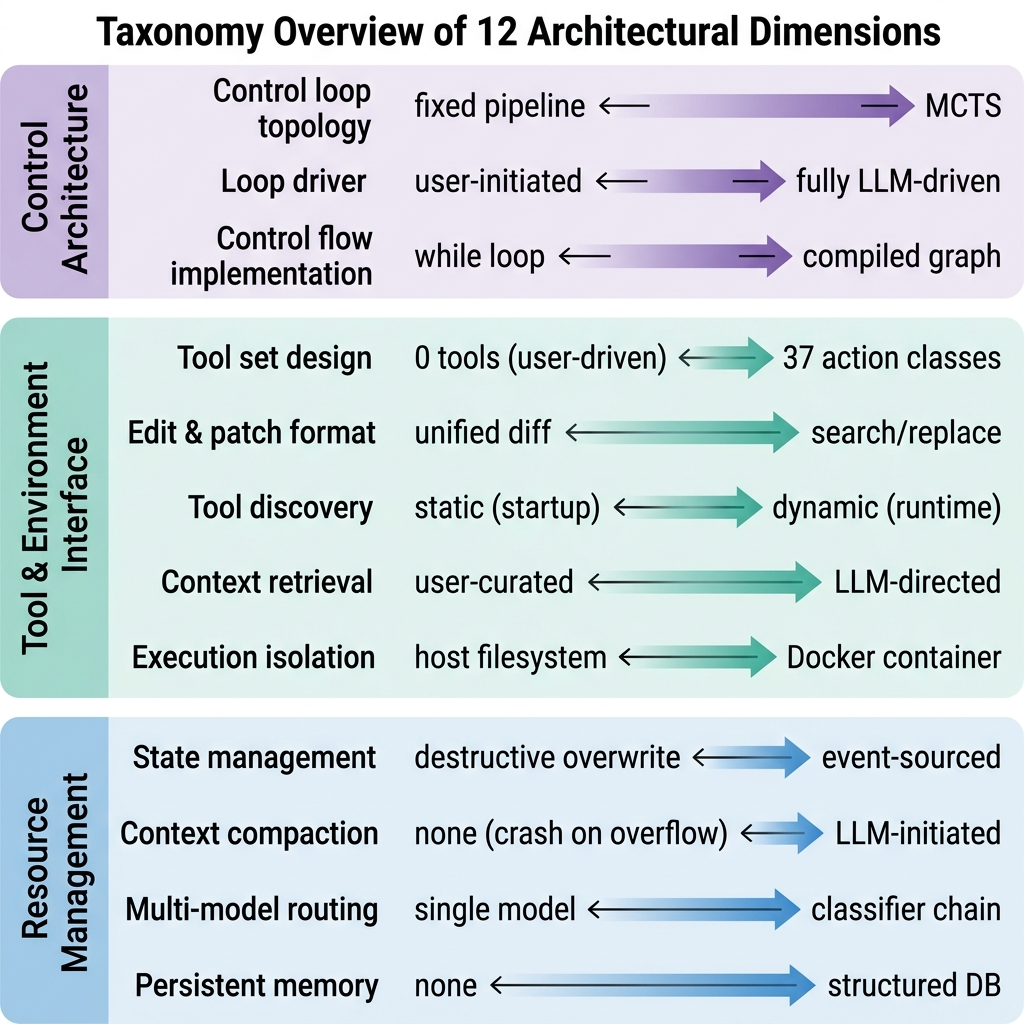}
\caption{Taxonomy overview: 12 dimensions organized into three architectural layers.
Each dimension is a continuous spectrum with observed endpoints from the corpus of
13 agents. Agents occupy positions along these spectra rather than falling into
discrete categories (Section~\ref{sec:discussion:spectra}).}
\label{fig:taxonomy-overview}
\end{figure*}

The analysis of the 13 open-source coding agent scaffolds reveals a taxonomy
organized into three layers: control architecture (how the agent decides what to
do next), tool and environment interface (how the agent interacts with code and
execution environments), and resource management (how the agent manages context,
state, and models). Within each layer, architectural choices fall along
continuous gradients rather than into discrete categories. In this section,
each dimension is presented as a range of observed strategies, agents are
placed along it with source-code evidence, and cross-cutting patterns are
identified. Figure~\ref{fig:taxonomy-overview} provides an overview of
the full taxonomy structure.

\subsection{Layer 1: Control Architecture}
\label{sec:results:control}

The control layer determines how an agent orchestrates its actions. Three
dimensions capture the key variation: (1)~the topology of the control loop,
(2)~what drives it, and (3)~how it is implemented in code.

\subsubsection{Control Loop Strategies}
\label{sec:results:loop-spectrum}

Control loops range from fixed pipelines with no feedback to
tree-structured search with backpropagation
(Table~\ref{tab:control-loop}), and these loop types are not mutually
exclusive; agents frequently nest one loop type inside another. This
composability means the table understates the actual control complexity: an agent classified as a
``phased loop'' may contain a full ReAct loop inside each phase.
\textsf{AutoCodeRover}, for example, runs a multi-turn LLM interaction
inside each stage of its pipeline
(\texttt{agent\_search.py:88--163}): a generator function yields tool
selections to the caller, which executes them and sends results back,
making it simultaneously a phased agent and an
iterative agent depending on the level of abstraction.
\textsf{Moatless Tools} takes composability further by decoupling the
inner agent from the outer control flow entirely: its
\texttt{ActionAgent} (a single-step executor: one LLM call producing
actions, then executing them) can be driven either by
\texttt{AgenticLoop} (which calls it repeatedly to produce ReAct
behavior) or by \texttt{SearchTree}
for MCTS-based exploration, with no changes to the agent code itself.
This separation means that the choice between
sequential and tree-search exploration is a configuration decision
rather than an architectural one.

\begin{table*}[htbp]
\caption{Control loop strategies. Agents ordered from least to most flexible
exploration strategy.}
\label{tab:control-loop}
\centering
\footnotesize
\begin{tabularx}{\textwidth}{l p{3.5cm} X}
\toprule
Position & Agents & Mechanism \\
\midrule
Fixed pipeline & \textsf{Agentless} &
  10-stage pipeline of independent scripts connected by JSONL files on
  disk. No feedback loop between stages. \\
User-driven loop & \textsf{Aider} &
  Outer loop is user-initiated (each edit cycle requires user input).
  Inner generate-test-repair loop is autonomous for up to
  \texttt{max\_reflections} iterations. \\
Sequential ReAct loop & \textsf{SWE-agent}, \textsf{OpenHands}, \textsf{Codex CLI}, \textsf{Gemini CLI},
  \textsf{mini-swe-agent}, \textsf{Cline}, \textsf{OpenCode} &
  Standard thought--tool--observation cycle. LLM selects the next action;
  loop terminates on completion signal or budget exhaustion. \\
Phased loop & \textsf{AutoCodeRover}, \textsf{Prometheus} &
  Distinct stages with different tool access. \textsf{AutoCodeRover}: search-then-patch
  phase separation. \textsf{Prometheus}: LangGraph~\citep{langgraph}
  state machine with explicit edges. \\
Depth-first tree search & \textsf{DARS-Agent} &
  ReAct steps form nodes in a search tree. At branch points,
  the environment is reset and replayed from root; an LLM
  critic selects among sampled alternatives. \\
Full MCTS & \textsf{Moatless Tools} &
  Select--Expand--Simulate--Backpropagate with reward values ($-100$ to $+100$)
  and visit counts. Pluggable selector interface; a discriminator selects the
  best finished trajectory. \\
\bottomrule
\end{tabularx}
\end{table*}

\paragraph{Tree search strategies.}
Three agents span a gradient from flat sampling to informed search,
illustrating increasingly sophisticated approaches to exploring the
space of possible solutions.

At the simplest end, \textsf{Agentless} uses independent sampling: after
localizing the relevant code, it prompts the LLM to generate
candidate patches independently (the default configuration generates
20; the original paper uses 4 location samples $\times$ 10 patches
each for ${\sim}$40).
Each generation sees the
same context but may produce a different fix. It then selects the
final patch by majority vote across these candidates. There is no tree structure and no
interaction between candidates; each patch is generated in isolation.

\textsf{DARS-Agent} introduces tree-structured search into its main
execution loop. Unlike \textsf{Agentless}, which separates localization
and repair into distinct phases, \textsf{DARS-Agent} has no phase
separation: search commands are available alongside edit and execute
commands throughout. The agent builds a search tree where
each node represents an action (such as editing a file, creating a
file, or submitting a patch). At each of these branch points, the agent
generates multiple alternative actions, then uses an LLM critic to
select among them: the critic is prompted with the alternatives and
responds with a \texttt{<best\_action\_index>} tag indicating its
choice. However, unlike classical
tree search, \textsf{DARS-Agent} has no numeric reward signals and no
backpropagation of results to earlier nodes; the critic makes greedy
local decisions without considering how earlier choices affected later
outcomes.

\textsf{Moatless Tools} implements full Monte Carlo Tree Search
(MCTS)~\citep{browne2012mcts}, the same algorithm used in game-playing
systems like AlphaGo~\citep{silver2016alphago}. Each node in the search tree receives numeric
reward values (ranging from $-100$ to $+100$), and the algorithm
maintains visit counts to balance exploration of untried paths against
exploitation of promising ones. After expanding a node, rewards are
backpropagated up the tree to update ancestor nodes, allowing the
search to learn from later outcomes and redirect effort toward more
promising branches (\texttt{search\_tree.py:326--345}).

The progression from flat sampling to full MCTS reveals a design
tradeoff: richer search strategies require a mechanism to manage
execution state across branches. \textsf{Moatless Tools} addresses this
with shadow-mode execution, where file modifications are tracked in
memory rather than written to disk; this allows the search to branch at any point
without costly filesystem operations. \textsf{DARS-Agent} takes a
different approach: at each branch point, it resets the Docker
environment to a clean state and replays all actions from the root of
the tree to reach the desired branch,
which is correct but expensive at deeper search depths.

The remaining CLI agents (\textsf{Cline}, \textsf{Codex CLI},
\textsf{Gemini CLI}, \textsf{OpenCode}) all implement standard sequential
ReAct loops without phased structure or tree search, placing them in the
same middle band as the other ReAct agents in Table~\ref{tab:control-loop}.
\textsf{Aider} is a special case: although it appears in this band, its
outer loop is user-driven rather than agentic
(Section~\ref{sec:results:loop-driver}),
with the LLM producing text-format edits rather than selecting tools.
However, \textsf{Aider}'s inner loop (\texttt{base\_coder.py:932}) is
autonomous: after the LLM produces edits, the scaffold runs linting and
tests, and if either fails, re-prompts the LLM with the error output
for up to \texttt{max\_reflections} iterations. This generate-test-repair
cycle is the only part of \textsf{Aider} where multi-turn LLM interaction
occurs without user input.

\subsubsection{Loop Driver}
\label{sec:results:loop-driver}

Orthogonal to loop topology is the question of what drives the loop
(Table~\ref{tab:loop-driver}), and this dimension is arguably the most
fundamental architectural distinction.
\textsf{Aider} sits at one extreme: the LLM never runs \textsf{grep}, never opens a
file it was not given, and never decides ``I should look at module Y.'' All
navigation responsibility falls on the user, with a PageRank-weighted repo
map providing passive context. At the other
extreme, 9 of 13 agents give the LLM full autonomy over tool selection,
with \textsf{Prometheus} occupying a hybrid position where the LLM drives
tool selection within each graph node but scaffold-controlled edges govern
transitions between nodes.
Between these poles, \textsf{Agentless} and \textsf{AutoCodeRover} occupy an intermediate
position where the scaffold sequences phases but the LLM makes decisions
within each phase.

\begin{table}[htbp]
\caption{Loop driver strategies. Who decides what happens next?}
\label{tab:loop-driver}
\centering
\footnotesize
\begin{tabularx}{\textwidth}{lX}
\toprule
Driver & Agents \\
\midrule
User-driven &
  \textsf{Aider}: the LLM has 0 callable tools.
  The user selects files (\texttt{/add}), runs searches, and provides
  context. The LLM produces edits in a text format parsed by the scaffold
  (\texttt{base\_coder.py:2296--2304}). \\
Scaffold-driven &
  \textsf{Agentless}, \textsf{AutoCodeRover}: the scaffold controls sequencing and calls
  the LLM at fixed points. In \textsf{Agentless}, each LLM call is single-turn
  with no conversation state. In \textsf{AutoCodeRover}, the scaffold manages
  phase transitions across up to four stages: optional reproducer
  generation, optional SBFL fault localization, search, and patch
  generation. \\
LLM-driven &
  \textsf{SWE-agent}, \textsf{OpenHands}, \textsf{Codex CLI}, \textsf{Gemini CLI}, \textsf{Cline}, \textsf{mini-swe-agent},
  \textsf{Moatless Tools}, \textsf{DARS-Agent}, \textsf{OpenCode}: the LLM selects tools and
  controls exploration. \textsf{Prometheus} is a hybrid: within each
  graph node, the LLM drives tool selection via ReAct, but transitions
  between nodes are scaffold-controlled (conditional edges on state
  fields such as \texttt{state["reproduced\_bug"]}). \\
\bottomrule
\end{tabularx}
\end{table}

The loop driver has implications beyond control flow. User-driven agents
sidestep the bug localization bottleneck identified in prior trajectory
analyses: if the user selects files, incorrect localization is a user error,
not an agent failure. LLM-driven agents must solve localization as part of
the task, making retrieval strategy (Section~\ref{sec:results:retrieval})
a critical design choice.

\subsubsection{Control Flow Implementation}
\label{sec:results:control-impl}

The semantic loop types above are implemented through four distinct
code-level mechanisms (Table~\ref{tab:control-impl}): (1)~imperative
while loops, where the scaffold calls the LLM in a \texttt{while True}
loop and breaks on a termination signal (8 of 13 agents);
(2)~recursion, where each tool-use turn triggers a recursive function
call (\textsf{Cline}); (3)~graph-as-control-flow, where a compiled
state machine defines transitions and cycles (\textsf{Prometheus}); and
(4)~exception-based signaling, where special exception types carry
control messages between the loop body and the outer handler
(\textsf{mini-swe-agent}).

\begin{table}[htbp]
\caption{Control flow implementation variants. This table classifies agents
by code-level mechanism, which is orthogonal to the semantic loop strategy in
Table~\ref{tab:control-loop}: an agent classified as ``user-driven'' by loop
strategy (e.g., \textsf{Aider}) may still use an imperative while loop at the
implementation level.}
\label{tab:control-impl}
\centering
\footnotesize
\begin{tabularx}{\textwidth}{lX}
\toprule
Implementation & Agents \\
\midrule
Imperative while loop &
  \textsf{SWE-agent}, \textsf{OpenHands}, \textsf{Codex CLI}, \textsf{Gemini CLI}, \textsf{mini-swe-agent},
  \textsf{Aider}, \textsf{OpenCode} \\
Fixed pipeline (no loop) &
  \textsf{Agentless}: sequential scripts; each LLM call is single-turn
  with no agent loop. The Anthropic path contains a bounded
  \texttt{for} loop (up to 10 iterations, \texttt{model.py:148--284}),
  but the primary architecture has no loop. \\
Recursion &
  \textsf{Cline}: \texttt{recursivelyMakeClineRequests}
  (\texttt{task/index.ts:2268}). The call stack grows linearly
  with conversation length. \\
Graph-as-control-flow &
  \textsf{Prometheus}: LangGraph compiled state machine with explicit edges
  (\texttt{issue\_graph.py:22--134}). Cycles in the graph create loops;
  recursion limits (30 for \texttt{IssueBugSubgraph}, 150 for
  \texttt{BugReproductionSubgraph}; dynamic formulas scale to hundreds
  of steps for multi-candidate runs) serve as the termination
  guarantee. \\
Exception-based signaling &
  \textsf{mini-swe-agent}: \texttt{InterruptAgentFlow} hierarchy
  (\texttt{Submitted}, \texttt{LimitsExceeded}, \texttt{FormatError})
  carries messages as payloads. The \texttt{run()} method catches these,
  injects messages into history, and either continues or breaks
  (\texttt{default.py:88--96}). \\
\bottomrule
\end{tabularx}
\end{table}

The graph-based approach is qualitatively different from the others:
control flow is inspectable, serializable, and checkpointable by the
framework. \textsf{Prometheus}'s subgraphs define their own state types,
and each subgraph is wrapped in a \texttt{SubgraphNode} class that
translates between the parent graph's state and the child graph's
state. For example,
the parent \texttt{IssueState} holds \texttt{issue\_title},
\texttt{issue\_body}, and \texttt{issue\_comments}; the
\texttt{IssueBugSubgraphNode} wrapper passes these as keyword
arguments to \texttt{IssueBugSubgraph.invoke()}. At least four levels of nesting
exist: \texttt{IssueGraph} $\to$ \texttt{IssueBugSubgraph} $\to$
\texttt{IssueVerifiedBugSubgraph} $\to$
\texttt{ContextRetrievalSubgraph}.

\textsf{Cline}'s recursive implementation is the only instance of
recursion for the main agent loop in the corpus. While semantically
equivalent to iteration, it means the JavaScript call stack grows with
each tool-use turn. In practice, Node.js default stack limits
(typically thousands of frames) are unlikely to be reached during
normal sessions, but the design has architectural consequences: each
recursive frame retains local state, making the control flow harder to
serialize or checkpoint compared to iterative approaches. Seven agents (\textsf{SWE-agent},
\textsf{OpenHands}, \textsf{Codex CLI}, \textsf{Gemini CLI},
\textsf{mini-swe-agent}, \textsf{Aider},
\textsf{OpenCode}) use imperative while loops.
\textsf{Agentless} has no agent loop at all (its pipeline is a fixed
sequence of scripts), though its Anthropic code path contains a small
iteration loop. \textsf{Codex CLI}'s while loop is notably more complex
than the others: it uses a two-level event-driven architecture with async
channels, though the logical pattern remains a standard ReAct loop. Within this group,
\textsf{OpenCode} is architecturally distinctive for layering a global
publish-subscribe event bus on top of its while loop
(\texttt{packages/opencode/src/bus/}). Components communicate via typed
events rather than direct function
calls. No other CLI agent
in the corpus uses an event bus for inter-component communication;
\textsf{OpenHands} uses event sourcing for state management
(Section~\ref{sec:results:state}), but its event stream serves a
different purpose (persistence and replay) than \textsf{OpenCode}'s
pub/sub bus (decoupled runtime communication).

\subsection{Layer 2: Tool and Environment Interface}
\label{sec:results:tools}

The interface layer captures how agents interact with code and execution
environments. Five dimensions characterize the design space: (1)~tool set
design, (2)~edit and patch format, (3)~tool discovery strategy,
(4)~context retrieval paradigm, and (5)~execution isolation.

\subsubsection{Tool Set Design}
\label{sec:results:toolset}

Tool counts range from 0 (\textsf{Aider}) to 37 action classes
(\textsf{Moatless Tools}), but raw counts obscure a convergence in
capability categories (Table~\ref{tab:tool-count}). Despite this range,
the same four core capability categories (read, search, edit, execute) appear
across all LLM-driven agents (as classified in
Section~\ref{sec:results:loop-driver}). A fifth category, validate
(dedicated test-running or linting tools), appears only in
\textsf{Moatless Tools}; other agents subsume validation under their
execute capability. The two scaffold-driven agents that use LLM-callable
tools (\textsf{AutoCodeRover}, \textsf{Agentless}) intentionally restrict
their tool sets to subsets of these categories, reflecting their phased
architectures. Agents with fewer tools achieve coverage through composition:
\textsf{mini-swe-agent}'s single \textsf{bash} tool
covers all four categories by
delegating to shell commands. Agents with more tools decompose these
categories into finer-grained operations (\textsf{Moatless Tools} has separate
\textsf{FindClass}, \textsf{FindFunction}, \textsf{FindCodeSnippet},
\textsf{SemanticSearch}, \textsf{GrepTool}, and \textsf{GlobTool}
actions for the search category alone). \textsf{OpenHands} stands out for
including a BrowserGym-based headless browser~\citep{chezelles2024browsergym} as a first-class tool,
the only agent in the corpus with a built-in
web browsing tool for navigating and interacting with web pages. It also provides a \textsf{think} tool for
logging reasoning without side effects and a
\textsf{request\_condensation} tool that lets the LLM request context
compaction (Section~\ref{sec:results:compaction}). \textsf{OpenCode} has
a comparable tool count (18 built-in) and includes distinctive tools not
found in other agents: experimental LSP integration for symbol navigation,
a \textsf{skill} meta-tool that loads
user-defined domain-specific instructions and workflows from the
filesystem, and a \textsf{batch} tool for grouping
multiple operations. \textsf{Codex CLI} (${\sim}$20+ tools) is notable
for including meta-tools that let the LLM discover additional tools at
runtime: \textsf{tool\_search} queries registered app tools, and
\textsf{tool\_suggest} recommends tools
for the current task. It also exposes a
\textsf{request\_permissions} tool that lets the LLM ask for elevated
sandbox access mid-session, making
the agent's own permissions a negotiable resource rather than a fixed
constraint.

\begin{table*}[htbp]
\caption{Tool set size and capability coverage. Tool count reflects
LLM-callable tools; user-facing commands (e.g., \textsf{Aider}'s ${\sim}$38 slash commands)
are excluded.}
\label{tab:tool-count}
\centering
\footnotesize
\begin{tabular}{rlcccccl}
\toprule
Agent & Tools & Read & Search & Edit & Execute & Validate & Notes \\
\midrule
\textsf{Aider} & 0 & & & \checkmark$^*$ & & & $^*$Text-parsed edits, 13 formats \\
\textsf{Agentless} & 0--1 & & & \checkmark$^*$ & & & $^*$1 simulated tool in Anthropic path \\
\textsf{AutoCodeRover} & 8 & \checkmark & \checkmark & & & & All search/read; no edit tools \\
\textsf{OpenHands} & 9+ & \checkmark & & \checkmark & \checkmark & & +MCP; search via \textsf{bash} \\
\textsf{Moatless Tools} & 15--37 & \checkmark & \checkmark & \checkmark & \checkmark & \checkmark & 37 classes; ${\sim}$15 typical per session \\
\textsf{SWE-agent} & 3--35 & \checkmark & \checkmark & \checkmark & \checkmark & & 3 default; 35 across 15 bundles \\
\textsf{DARS-Agent} & ${\sim}$15 & \checkmark & \checkmark & \checkmark & \checkmark & & Inherited from \textsf{SWE-agent} fork \\
\textsf{Codex CLI} & ${\sim}$20+ & \checkmark & \checkmark & \checkmark & \checkmark & & +MCP; +sub-agent spawning \\
\textsf{Gemini CLI} & 17+ & \checkmark & \checkmark & \checkmark & \checkmark & & +MCP; +tool discovery subprocess \\
\textsf{Prometheus} & 17 & \checkmark & \checkmark & \checkmark & \checkmark & & 1--12 bound per graph node \\
\textsf{OpenCode} & 18+ & \checkmark & \checkmark & \checkmark & \checkmark & & +MCP; +plugins; +custom tools \\
\textsf{Cline} & 27+ & \checkmark & \checkmark & \checkmark & \checkmark & & +MCP; largest flat built-in set \\
\textsf{mini-swe-agent} & 1 & & & & \checkmark & & All capabilities via \textsf{bash} \\
\bottomrule
\end{tabular}
\end{table*}

\paragraph{Per-node tool scoping.}
\textsf{Prometheus} is the only agent that binds different tool subsets to different
decision points. Its \texttt{EditNode} sees 5 tools (file operations),
while its \texttt{BugReproducingWriteNode} sees only \textsf{read\_file},
and its \texttt{BugFixVerifyNode} sees only \textsf{run\_command}.
This structural guardrailing
constrains the action space at each step. \textsf{AutoCodeRover} achieves a similar
effect through phase separation: the patch agent is not given search
tools and is instead prompted to produce patches directly from the
context gathered during the search phase. This is a workflow-level
constraint rather than a configurable binding.

\paragraph{Search-only tools in \textsf{AutoCodeRover}'s localization phase.}
\textsf{AutoCodeRover}'s search agent has 8 LLM-callable tools, all of
which are read-only search operations;
the search LLM cannot edit
files, run commands, or modify state. This is the most constrained
tool set of any agent phase that uses LLM-callable tools.
\textsf{AutoCodeRover} does generate patches, but this happens in a
separate patch agent phase where the
LLM is prompted to produce a patch directly in its output, without
calling any tools. The philosophy is that localization and repair are
distinct tasks with different tool requirements: the search phase
needs structured code navigation, while the patch phase needs only the
context gathered during search and the LLM's code generation
capability.

\paragraph{Proxy agent for tool invocation in \textsf{AutoCodeRover}.}
\textsf{AutoCodeRover} uses a secondary LLM call
(\texttt{agent\_proxy.py:81--82}) to convert the search agent's
natural-language tool selections into structured JSON. The search agent ``thinks out loud''
about which tools to call; the proxy agent extracts the structured
invocation with up to 5 retries. This
adds one LLM call per search round (with 15 rounds possible, that is
up to 15 extra LLM calls) but avoids requiring the search agent to produce
structured output directly. No other agent uses a secondary LLM call for
tool call parsing; all others rely on function calling APIs, regex parsing,
or XML extraction.

\subsubsection{Edit and Patch Format}
\label{sec:results:edit-format}

How agents translate LLM output into code changes reveals a convergence
toward a shared interface (Table~\ref{tab:edit-format}). The
\textsf{str\_replace\_editor} tool, which takes \textsf{old\_str} and
\textsf{new\_str} arguments for exact string replacement, appears in 5
of 13 agents (\textsf{OpenHands}, \textsf{SWE-agent},
\textsf{Codex CLI}'s \textsf{apply\_patch} in freeform mode,
\textsf{Agentless} as one of three repair output formats,
and \textsf{Moatless Tools} as \textsf{StringReplace}). This
convergence is notable because these agents were developed independently;
the shared interface reflects a common discovery that exact string matching
is more reliable than line-number-based or unified-diff-based editing for
LLM-generated patches~\citep{yang2024sweagent}.

\begin{table}[htbp]
\caption{Edit/patch format variants.}
\label{tab:edit-format}
\centering
\footnotesize
\begin{tabularx}{\textwidth}{lX}
\toprule
Format & Agents \\
\midrule
String replacement (function calling) &
  \textsf{OpenHands} and \textsf{SWE-agent}
  (\textsf{str\_replace\_editor}),
  \textsf{Codex CLI} (\textsf{apply\_patch}),
  \textsf{OpenCode} (\textsf{edit}) \\
Write tool (function calling) &
  \textsf{Gemini CLI}, \textsf{Cline} \\
Text-parsed edit blocks &
  \textsf{Aider} (13 formats), \textsf{DARS-Agent} \\
Simulated tool use &
  \textsf{Agentless} (Anthropic path) \\
Pydantic-schema actions &
  \textsf{Moatless Tools}, \textsf{Prometheus} \\
\bottomrule
\end{tabularx}
\end{table}

On the parsing side, \textsf{SWE-agent} supports 10 different output
parsers (\texttt{FunctionCallingParser}, \texttt{ThoughtActionParser},
\texttt{XMLThoughtActionParser}, and others), enabling the same tool
set to work across models with different output conventions. Together
with \textsf{Aider}'s 13 model-specific edit formats, this represents
the highest degree of model-agnosticism in output interfacing observed
in the corpus; however, the two approaches address different layers
(\textsf{SWE-agent} adapts tool \emph{parsing}, while \textsf{Aider}
adapts the \emph{edit format} itself).

\textsf{Aider} is an outlier with 13 registered edit formats, each implemented as
a separate coder subclass. The format
is selected based on model capabilities: some models handle unified diffs
better, others work better with SEARCH/REPLACE blocks. This treatment of
edit format as a first-class, model-specific architectural component is
unique in the corpus, though \textsf{OpenCode} implements a simpler
version of the same idea: its tool registry selects between
\textsf{edit} (string replacement) and \textsf{apply\_patch} (unified
diff) based on model capabilities,
offering two formats rather than thirteen.

\textsf{Agentless}' ``simulated tool use'' in its Anthropic path is
architecturally unique. When using Anthropic's API, the LLM calls a
\textsf{str\_replace\_editor} tool, but every call receives the same
hardcoded response: ``File is successfully edited''
regardless of input. The LLM never sees actual
file state after its edits; the tool calls are extracted and applied
post-hoc. This uses the
tool-calling API as a structured output extraction technique rather than
for actual execution.

The remaining agents use variants of the same approach.
\textsf{AutoCodeRover} uses custom XML-like
\texttt{<original>}/\texttt{<patched>} tags,
while \textsf{Gemini CLI} and
\textsf{Cline} combine whole-file write tools with search-and-replace
tools functionally similar to \textsf{str\_replace\_editor}.
\textsf{mini-swe-agent} is the only agent that edits files directly via
shell commands rather than producing patches for the scaffold to apply.

\subsubsection{Tool Discovery Strategy}
\label{sec:results:tool-discovery}

Tool discovery ranges from fully static (tools fixed at initialization) to
genuinely dynamic (tools added during a session)
(Table~\ref{tab:tool-discovery}). Nearly half of the agents (6 of 13) use
fully static tool registration: all tools are defined at initialization
and remain unchanged throughout the session. This is the default approach for benchmark agents
(\textsf{Agentless}, \textsf{AutoCodeRover}, \textsf{mini-swe-agent},
\textsf{DARS-Agent}, \textsf{Moatless Tools}) and for \textsf{Aider},
where the LLM has no callable tools at all. \textsf{SWE-agent} and
\textsf{OpenHands} occupy a middle ground: their tool sets are
config-conditional (different tool bundles load depending on
configuration), but once loaded, the set is fixed for a given attempt.
(In \textsf{SWE-agent}'s retry loop, different attempts can load
different tool bundles if the \texttt{agent\_configs} specify different
configurations.)

\begin{table}[htbp]
\caption{Tool discovery strategies.}
\label{tab:tool-discovery}
\centering
\footnotesize
\begin{tabularx}{\textwidth}{p{2.5cm}X}
\toprule
Strategy & Agents \\
\midrule
Static (all at init) &
  \textsf{Aider}, \textsf{Agentless}, \textsf{AutoCodeRover}, \textsf{mini-swe-agent}, \textsf{DARS-Agent},
  \textsf{Moatless Tools} \\
Per-phase scoping (static per node) &
  \textsf{Prometheus}, \textsf{AutoCodeRover} \\
Config-conditional &
  \textsf{SWE-agent}, \textsf{OpenHands} \\
Per-turn dynamic rebuild &
  \textsf{Codex CLI} (\texttt{built\_tools()} called every sampling request,
  \texttt{codex.rs:6156--6164}) \\
Dynamic (MCP + subprocess) &
  \textsf{Gemini CLI} (tool discovery subprocess, \texttt{tool-registry.ts:312--439}),
  \textsf{Cline} (MCP servers connected/disconnected at runtime),
  \textsf{OpenCode} (static core tools + dynamic custom tools from config dirs,
  plugin system, and MCP integration) \\
\bottomrule
\end{tabularx}
\end{table}

At the dynamic end, \textsf{Codex CLI}'s per-turn tool rebuild is
distinctive: while most agents construct the tool set once (or once per
phase), \textsf{Codex CLI} calls \texttt{built\_tools()} for every LLM
sampling request, incorporating any MCP
server changes, newly enabled connectors, or dynamic tools added during
the session. The tool set is, in principle, different at every LLM call.
\textsf{Gemini CLI}'s \texttt{toolDiscoveryCommand} feature
takes a different approach: an
external subprocess outputs \texttt{FunctionDeclaration[]} as JSON,
allowing arbitrary tool sources without modifying the agent code.
\textsf{Cline} and \textsf{OpenCode} similarly support runtime tool
changes through MCP server connections and plugin systems.
\textsf{Prometheus} and \textsf{AutoCodeRover} fall between static and
dynamic: their tool sets are fixed within each phase or graph node, but
different phases expose different tool subsets (as discussed in
Section~\ref{sec:results:toolset}).

\subsubsection{Context Retrieval Paradigm}
\label{sec:results:retrieval}

Context retrieval (how the agent finds relevant code) shows substantial
architectural variation, with seven distinct strategy types ranging
from simple keyword search to knowledge graph traversal
(Table~\ref{tab:retrieval}). The 13 agents divide into two retrieval
paradigms that reflect fundamentally different assumptions about where
code understanding should live.

\begin{table*}[htbp]
\caption{Context retrieval strategies. Agents may use multiple strategies.}
\label{tab:retrieval}
\centering
\footnotesize
\begin{tabularx}{\textwidth}{p{2.2cm}p{3cm}X}
\toprule
Strategy & Agents & Mechanism \\
\midrule
Keyword/regex search &
  \textsf{SWE-agent}, \textsf{OpenHands}, \textsf{Codex CLI}, \textsf{Gemini CLI}, \textsf{Cline}, \textsf{mini-swe-agent},
  \textsf{DARS-Agent}, \textsf{OpenCode} &
  LLM invokes \textsf{grep}, \textsf{find}, \textsf{ripgrep} as tools. \\
Repo map (static analysis) &
  \textsf{Aider} &
  PageRank-weighted tree-sitter tag index. Builds a NetworkX dependency
  graph; identifiers mentioned in conversation get a 10x boost; chat
  files get a 50x boost (\texttt{repomap.py:365--574}). \\
AST-aware search &
  \textsf{AutoCodeRover}, \textsf{Moatless Tools}, \textsf{DARS-Agent}, \textsf{Prometheus} &
  \textsf{search\_class}, \textsf{search\_method}: structure-aware queries
  over pre-built AST indices. \textsf{AutoCodeRover}'s AST tools are
  Python-only; \textsf{Prometheus}'s tree-sitter-based tools cover
  20 languages. \\
Knowledge graph traversal &
  \textsf{Prometheus} &
  Neo4j graph built from tree-sitter ASTs covering 20 languages.
  11 tools (10 graph traversal plus \texttt{read\_file}) query
  \texttt{FileNode}, \texttt{ASTNode},
  \texttt{TextNode} entities (\texttt{graph\_traversal.py:93--586}). \\
Embedding-based semantic search &
  \textsf{Moatless Tools} &
  FAISS~\citep{johnson2021faiss} vector store via LlamaIndex~\citep{llamaindex} (\texttt{code\_index.py:57}).
  The only agent with embedding-based retrieval as an LLM-callable tool. \\
Hierarchical localization &
  \textsf{Agentless} &
  File $\to$ class/function $\to$ line-level narrowing across
  pipeline stages. Each level sees only what the previous level
  identified (\texttt{FL.py:313--681}). \\
Classical fault localization &
  \textsf{AutoCodeRover} &
  SBFL with Ochiai scoring (\texttt{analysis/sbfl.py}).
  Unique in the corpus. \\
\bottomrule
\end{tabularx}
\end{table*}

The first paradigm, which eight agents adopt (\textsf{SWE-agent},
\textsf{OpenHands}, \textsf{Codex CLI}, \textsf{Gemini CLI},
\textsf{Cline}, \textsf{mini-swe-agent}, \textsf{DARS-Agent},
\textsf{OpenCode}), treats the LLM as a
navigator: the agent provides general-purpose shell tools
(\textsf{grep}, \textsf{find}, \textsf{ripgrep}) and relies on the
LLM to formulate queries, interpret results, and decide where to look
next. The scaffold provides no code understanding of its own; it
simply executes the commands the LLM requests and returns the output.

The second paradigm invests in scaffold-side code understanding,
building structured representations of the codebase before or during
the task. \textsf{Aider} constructs a dependency graph from parsed
source code and uses PageRank to rank file relevance
(discussed below). \textsf{Agentless} narrows from files to classes to lines across
successive LLM calls, each seeing only what the previous stage
identified. \textsf{AutoCodeRover} and \textsf{Moatless Tools} parse
source code into abstract syntax trees (ASTs), enabling
structure-aware queries like ``find all methods named
\texttt{process}'' rather than text-pattern matching.
\textsf{Prometheus} goes furthest, constructing a Neo4j~\citep{neo4j} graph database
from ASTs covering 20 languages and providing 11 tools (10 graph
traversal plus \texttt{read\_file}) for querying relationships between code entities.

The paradigm choice correlates with loop driver
(Section~\ref{sec:results:loop-driver}): scaffold-driven agents tend
to invest in retrieval infrastructure, while LLM-driven agents tend to
trust the LLM to navigate with general-purpose tools. This correlation
is not surprising: if the scaffold controls sequencing, it has the
opportunity (and the need) to pre-process the repository; if the LLM
controls exploration, it can issue search commands on demand.

\paragraph{\textsf{Aider}'s PageRank repo map.}
\textsf{Aider}'s retrieval mechanism applies Google's PageRank
algorithm~\citep{brin1998pagerank} to source code structure.
Tree-sitter~\citep{treesitter} parses every file to extract symbol definitions and
cross-file references, forming a directed dependency graph
(\texttt{repomap.py:365--574}). PageRank scores rank files by structural
centrality, with context-aware boosts: identifiers mentioned in
conversation receive a 10x weight, files added to the chat receive 50x,
and a binary search fills the token budget in rank order. No other agent in the corpus uses
graph-theoretic relevance ranking. The approach trades startup cost
(parsing the full repository) for context quality: rather than relying
on the LLM to navigate the codebase, \textsf{Aider} precomputes
relevance from the repository's actual dependency structure.

\paragraph{\textsf{AutoCodeRover}'s SBFL.}
\textsf{AutoCodeRover} is the only agent that incorporates classical
fault localization~\citep{wong2016survey}. When failing tests are
available, it runs spectrum-based fault localization (SBFL) with Ochiai
suspiciousness scores~\citep{abreu2007accuracy}: methods executed
frequently by failing tests and rarely by passing tests rank highest
(\texttt{analysis/sbfl.py}). The top 5 suspicious methods
are presented to the LLM as advisory input,
supplementing its search-based exploration. This bridges two otherwise
disconnected communities (automated fault localization and LLM-based
repair)~\citep{zhang2023apr}; no other agent in the corpus uses test-execution-based
localization.

\subsubsection{Execution Isolation}
\label{sec:results:isolation}

Execution isolation spans from no sandboxing to speculative in-memory
execution (Table~\ref{tab:isolation}). For tree-search agents, execution
isolation interacts directly with search cost. \textsf{DARS-Agent}'s environment reset
performs a full Docker reset and
replays all actions from root to the branch point. At depth $N$, this
requires re-executing $N$ commands. DARS addresses this cost by limiting
branching to specific action types (edit, create, append, submit) via
\texttt{action\_expansion\_limit} (\texttt{default\_dars.yaml:179--184}).
\textsf{Moatless Tools} avoids replay entirely: its shadow mode
tracks file modifications in a \texttt{FileContext}
object without writing to disk, and each node clones its parent's file
context. This enables branching at any step without
environment cost.

\begin{table}[htbp]
\caption{Execution isolation strategies.}
\label{tab:isolation}
\centering
\footnotesize
\begin{tabularx}{\textwidth}{lX}
\toprule
Isolation level & Agents \\
\midrule
None (local shell) &
  \textsf{Gemini CLI}, \textsf{Aider}, \textsf{OpenCode} \\
Stateless subshells &
  \textsf{mini-swe-agent} (\texttt{subprocess.run()},
  \texttt{local.py:28--39}) \\
Platform sandboxing &
  \textsf{Codex CLI} (Bubblewrap+Landlock on Linux,
  Seatbelt on macOS) \\
Docker container &
  \textsf{SWE-agent}, \textsf{OpenHands}, \textsf{DARS-Agent}, \textsf{AutoCodeRover},
  \textsf{Prometheus} \\
Shadow git checkpoints &
  \textsf{Cline} (isolated git repo tracks state per
  tool execution, \texttt{CheckpointTracker.ts}) \\
Shadow mode (in-memory) &
  \textsf{Moatless Tools} (\texttt{shadow\_mode} flag,
  \texttt{agent.py:57}) \\
Not applicable &
  \textsf{Agentless}: LLM never executes arbitrary commands;
  Docker used only for test execution via SWE-bench harness \\
\bottomrule
\end{tabularx}
\end{table}

\paragraph{\textsf{Cline}'s shadow git.}
\textsf{Cline}'s \texttt{CheckpointTracker} creates an
isolated git repository per workspace that records file system state after
each tool execution. This enables diff-based rollback without touching the
user's real git history, and without requiring Docker. It handles nested
git repos by temporarily disabling them and validates against sensitive
directories. This is a middle ground between no isolation and
containerization, specific to the IDE context where \textsf{Cline} operates.

\paragraph{Safety approaches diverge.}
Agents without container isolation pursue strikingly different safety
strategies, suggesting the ecosystem has not converged on an isolation
standard for interactive agents.

\textsf{Gemini CLI} uses a rule-based policy engine that assigns per-tool approval
requirements. Before executing a tool call, the engine checks whether
the tool requires user confirmation, is automatically allowed, or is
blocked entirely. The rules are configurable, letting users adjust the
safety boundary for their workflow.

\textsf{Codex CLI} combines two mechanisms. At the OS level, it uses
platform-specific sandboxing (Bubblewrap~\citep{bubblewrap} and Landlock~\citep{landlock} on Linux,
Seatbelt on macOS) to restrict filesystem and network access. On top
of this, a Guardian safety subagent, a separate LLM
(\texttt{gpt-5.4})\footnote{Model identifiers such as \texttt{gpt-5.4}, \texttt{gpt-5.1-codex-mini}, and \texttt{gpt-5.3-codex} reflect the strings observed in the analyzed commit (Appendix~\ref{sec:appendix:commits}). These names may differ in later versions of \textsf{Codex CLI}.}, evaluates each tool call's risk with structured
scoring on a 0--100 scale, blocking calls above a threshold of 80
(\texttt{guardian.rs}). This is the only agent in the corpus that uses
an LLM to evaluate the safety of another LLM's actions.

\textsf{Aider} takes the simplest approach: it relies on the user
being present and supervising. Since the user manually selects which
files to edit and reviews all proposed changes before they are
applied, the human serves as the safety boundary.

\textsf{Cline} provides the most granular approval system in the corpus:
per-tool and per-scope approval settings, with a
\texttt{CommandPermissionController} that blocks dangerous shell
operators. Settings range from full autonomous mode (``YOLO'') to
per-command-pattern approval, giving users fine-grained control over the
trust boundary. \textsf{OpenCode}
uses a policy-based permission system similar to \textsf{Gemini CLI}'s:
tools can request runtime user approval with three options (Allow Once,
Always Allow, Reject) via a permission callback in individual tool
implementations, but provides no
OS-level sandboxing.
\textsf{mini-swe-agent} uses stateless subshells,
providing
process-level isolation without containerization.

The five Docker-based agents (\textsf{SWE-agent}, \textsf{OpenHands},
\textsf{DARS-Agent}, \textsf{AutoCodeRover}, \textsf{Prometheus})
rely on container boundaries as their primary safety mechanism: all commands
execute inside a container, and the host filesystem is never directly
exposed. \textsf{OpenHands}' Docker implementation is architecturally
distinctive: a FastAPI action server runs \emph{inside} the container,
and the host-side agent controller communicates with it via HTTP.
This creates a clean
API boundary between the agent and the execution environment, unlike
other Docker agents that exec commands directly into containers.
\textsf{Agentless}
occupies a unique position: because the LLM never executes arbitrary
commands (it only generates text that the scaffold parses), the
isolation requirements are fundamentally different from agentic
approaches. Docker is used only for test execution via the SWE-bench
harness, not for agent safety.

The spread from human supervision to LLM-based safety evaluation
reflects an open design question: as agents gain more autonomy
(Section~\ref{sec:results:loop-driver}), the safety mechanism must
scale correspondingly, but no standard approach has emerged.

\subsection{Layer 3: Resource Management}
\label{sec:results:resources}

The resource management layer addresses how agents handle the constraints
of finite context windows, API costs, and session boundaries. Four
dimensions capture the variation: (1)~state management, (2)~context
compaction, (3)~multi-model routing, and (4)~persistent memory.

\subsubsection{State Management}
\label{sec:results:state}

How agents represent and maintain conversation state ranges from simple
destructive overwrite to full event sourcing~\citep{fowler2005eventsourcing}
(Table~\ref{tab:state-mgmt}). At one extreme, \textsf{Aider}'s two-list
design (\texttt{cur\_messages} and \texttt{done\_messages}) is simple
but destructive: summarization replaces the contents of
\texttt{done\_messages}. At the other extreme, \textsf{OpenHands}'
event-sourced architecture stores immutable events and computes views
via the \texttt{View} class;
condensation inserts markers rather than deleting events, preserving
the full audit trail. Between these extremes, \textsf{OpenCode}'s
SQLite-backed hierarchy uses append-only
messages with 12 typed part variants, the most granular state
representation in the corpus after \textsf{OpenHands} and the only one
backed by a relational database rather than in-memory structures.
\textsf{Agentless} sits at the opposite end of the spectrum: it has no
conversation state at all, since each LLM call is single-turn. State
between pipeline stages is represented as immutable JSONL files on disk,
one JSON object per problem instance per line. This
file-based state bus makes the pipeline trivially resumable
(\texttt{--skip\_existing} checks instance presence in output) and
parallelizable, but means there is no conversation history to manage
because there is no conversation. \textsf{Codex CLI} stands out among the flat-list agents by adding dual
persistence: an append-only JSONL rollout file for human-readable replay
alongside a SQLite database for queryable
state and session resumption. It is also the
only flat-list agent that supports undo (\texttt{Op::Undo}) and thread
rollback (\texttt{Op::ThreadRollback}). \textsf{Cline} maintains two
parallel message lists, one for the LLM API and one for the UI,
synchronized via mutex-protected concurrent access; this dual-list
architecture reflects its IDE integration, where the UI needs a richer
representation than the API.

\begin{table}[htbp]
\caption{State management strategies.}
\label{tab:state-mgmt}
\centering
\footnotesize
\begin{tabularx}{\textwidth}{lX}
\toprule
Strategy & Agents \\
\midrule
Destructive &
  \textsf{Aider}: summarization overwrites
  \texttt{done\_messages}
  (\texttt{base\_coder.py:1024--1034}) \\
Flat list, preserved &
  \textsf{SWE-agent}, \textsf{Codex CLI}, \textsf{Gemini CLI}:
  raw history kept; filtered views created for LLM.
  \textsf{mini-swe-agent}: raw history kept with no filtering
  (messages sent to the LLM as-is) \\
Typed event log &
  \textsf{OpenCode}: SQLite-backed message/part hierarchy
  with 12 part types (Drizzle ORM). Append-only
  messages with mutable part states
  (\texttt{message-v2.ts}) \\
Graph-scoped &
  \textsf{Prometheus}: separate message lists per graph node
  (\texttt{edit\_messages}, \texttt{analyzer\_messages},
  \texttt{context\_provider\_messages}), reset at
  retry boundaries \\
Tree-structured (MCTS) &
  \textsf{Moatless Tools}: nodes in a tree with per-node file context
  snapshots, visit counts, and reward values \\
Tree-structured (greedy) &
  \textsf{DARS-Agent}: nodes in a tree with per-node expansion
  candidates and critic responses
  (\texttt{dars\_agent.py:294--297}). No MCTS statistics; branching
  uses greedy LLM critic selection, and state is recovered via
  Docker reset and action replay rather than file context snapshots. \\
Event-sourced &
  \textsf{OpenHands}: immutable \texttt{EventStream}; views computed
  from condensation markers
  (\texttt{memory/view.py:13--96}) \\
\bottomrule
\end{tabularx}
\end{table}

\textsf{Prometheus}'s graph-scoped state is structurally unique. Rather than a
single growing conversation, each LLM node in the graph maintains its own
message list. \texttt{ResetMessagesNode} clears specific message lists
when the graph cycles back for retries,
preventing unbounded accumulation
without any token-counting logic.

Tree-structured state stores per-node metadata that flat lists cannot
represent, but the two tree-search agents use it differently.
\textsf{Moatless Tools} maintains MCTS statistics (visit counts and
reward values) and clones file context
snapshots at each branch point.
\textsf{DARS-Agent} stores expansion candidates and critic responses
but has no MCTS statistics: it uses
greedy LLM critic selection and recovers branch state by resetting the
Docker environment and replaying actions from the root, rather than
maintaining per-node file snapshots.

\subsubsection{Context Compaction}
\label{sec:results:compaction}

All agents face the constraint of finite context windows, and compaction
strategies range from no management to LLM-initiated compaction
(Table~\ref{tab:compaction}).

\begin{table*}[!htbp]
\caption{Context compaction strategies.}
\label{tab:compaction}
\centering
\footnotesize
\begin{tabularx}{\textwidth}{p{2cm}p{2.3cm}X}
\toprule
Strategy & Agents & Mechanism \\
\midrule
None (crash on overflow) &
  \textsf{mini-swe-agent} &
  Unbounded growth; agent crashes on \texttt{ContextWindowExceededError}. \\
Rule-based truncation &
  \textsf{SWE-agent}, \textsf{DARS-Agent} &
  \textsf{SWE-agent}: 7 composable processors; keep first + last $N$ observations,
  elide rest. Polling parameter for prompt-cache preservation.
  \textsf{DARS-Agent}: adds \texttt{Last5Observations} in its fork. \\
Structural isolation &
  \textsf{Prometheus}, \textsf{AutoCodeRover} &
  \textsf{Prometheus}: per-node message scoping + resets at retry boundaries.
  \textsf{AutoCodeRover}: round limits (15) + result truncation. \\
Token-based selective inclusion &
  \textsf{Moatless Tools} &
  Greedy recent-first selection within token budget; per-observation
  \texttt{summary} fallback. \\
LLM summarization (scaffold-triggered) &
  \textsf{Aider}, \textsf{OpenHands}, \textsf{Gemini CLI}, \textsf{Codex CLI}, \textsf{OpenCode} &
  Automatic at token threshold. \textsf{Aider}: recursive hierarchical
  summarization. \textsf{Codex CLI}: pre-turn and mid-turn
  modes. \textsf{OpenCode}: two-phase
  (prune old outputs, then LLM summarization). \\
LLM summarization + verification &
  \textsf{Gemini CLI} &
  Summarize, then ``Probe'' verification turn to check for information loss. \\
LLM-initiated compaction &
  \textsf{Cline} &
  \textsf{condense} tool: LLM decides when to compact. Also supports
  scaffold-triggered compaction at token threshold. \\
\bottomrule
\end{tabularx}
\end{table*}

The corpus reveals two distinct
philosophies. ``Prevention'' agents bound
context growth structurally: \textsf{Prometheus} scopes messages per graph node and
resets them at retry boundaries; \textsf{AutoCodeRover} caps search rounds at 15
and shows at most 3 full results per query; \textsf{Moatless Tools} limits
trajectory depth through tree structure. ``Cure'' agents let context grow
and compress when needed: \textsf{Aider}, \textsf{OpenHands}, \textsf{Gemini CLI}, and \textsf{Codex CLI} all
trigger LLM-based summarization at token thresholds. The prevention
approach avoids summarization cost and information loss but requires the
scaffold to anticipate context growth patterns. \textsf{Agentless}
sidesteps the compaction problem entirely: because each LLM call is
single-turn with no conversation history, ``compaction'' reduces to
fitting code into a single prompt. When prompts exceed the 128K token
limit, \textsf{Agentless} progressively drops files from the end of the
ranked list and compresses function bodies to
signatures via \texttt{libcst}. This
pre-computation approach eliminates the need for runtime compaction by
never accumulating conversational state. Among the ``cure'' agents,
\textsf{OpenCode}'s two-phase strategy is the most surgical: it first
prunes verbose old tool outputs, maintaining message structure but
replacing outputs older than the most recent 40,000 tokens with truncation markers.
Only then does it trigger LLM-based
summarization via a dedicated compaction agent that can use a
cheaper model. This preserves more
conversational context than pure truncation while being more targeted
than full summarization. \textsf{Codex CLI} distinguishes between
pre-turn and mid-turn compaction: pre-turn
compaction runs before each user turn and clears reference context so
the next turn reinjects initial context cleanly; mid-turn compaction
runs when the token limit is hit during tool execution and injects
initial context above the last user message, matching the model's
training expectations about where context appears. This awareness of
compaction timing relative to the conversation structure is unique in the
corpus.

\paragraph{\textsf{SWE-agent}'s polling parameter.}
\textsf{SWE-agent}'s \texttt{LastNObservations} processor
includes a \texttt{polling} parameter that reveals a subtle interaction between compaction and API cost.
Many LLM providers offer prompt caching, where consecutive calls sharing
the same message prefix skip reprocessing. Without polling, every new
observation changes which messages are included, invalidating the cache.
With polling, truncation changes occur at a rate of only
$1/\texttt{polling}$ per step, keeping the prefix stable across
consecutive calls. This cost optimization operates at the intersection
of two otherwise independent concerns and has not been documented in
prior analyses of \textsf{SWE-agent}.

\paragraph{\textsf{Gemini CLI}'s verification probe.}
\textsf{Gemini CLI} is the only agent that validates its own context compaction. After
LLM summarization, it runs a ``Probe'' turn where the model checks whether
critical information was lost. This
self-correction mechanism addresses a known failure mode of LLM
summarization (lossy compression of technical details) at the cost of an
additional LLM call per compaction event.

\paragraph{\textsf{Cline}'s LLM-initiated compaction.}
\textsf{Cline}'s \textsf{condense} tool gives the LLM agency
over its own context management. The LLM can proactively request
summarization if it judges the context to be unwieldy. No other agent in
the corpus delegates the compaction decision to the LLM. \textsf{OpenHands}
provides a \texttt{CondensationRequestAction} that is conceptually similar,
but compaction can also be triggered automatically when history exceeds
condenser thresholds.
\textsf{OpenHands}' condenser architecture is the most extensible in the
corpus: nine pluggable implementations composable into pipelines via a
registry pattern. Because condensation
operates at the view level (inserting \texttt{CondensationAction} markers
rather than deleting events), the raw event stream is never modified,
enabling session replay even after aggressive compaction.

\subsubsection{Multi-Model Routing}
\label{sec:results:routing}

Multi-model routing (using different LLMs for different subtasks) ranges
from single-model simplicity to multi-strategy classifier chains
(Table~\ref{tab:model-routing}). Two agents use a single model
throughout: \textsf{mini-swe-agent} and \textsf{Agentless}.
\textsf{OpenHands} provides extensible routing infrastructure
but defaults to single-model
operation; its only implemented router makes narrow decisions based on
image presence and token limits. The remaining 10 agents route to
multiple models, but for different reasons.

\begin{table*}[htbp]
\caption{Multi-model routing strategies.}
\label{tab:model-routing}
\centering
\footnotesize
\begin{tabularx}{\textwidth}{p{2cm}p{2.3cm}X}
\toprule
Strategy & Agents & Mechanism \\
\midrule
Single model &
  \textsf{mini-swe-agent}, \textsf{Agentless} (per path) &
  One model throughout. \\
Router abstraction (default single) &
  \textsf{OpenHands} &
  \texttt{RouterLLM} base with \texttt{MultimodalRouter}: routes by
  image presence and token limits
  (\texttt{llm/router/rule\_based/impl.py:16--81}). Default: single
  model via \texttt{noop\_router}. \\
Role-based &
  \textsf{Aider}, \textsf{OpenCode} &
  Main/weak/editor for different subtasks. \textsf{Aider}: weak model for
  summarization and commit messages (\texttt{models.py:607--608}); editor model for architect mode
  (\texttt{architect\_coder.py:22--25}). \textsf{OpenCode}: per-agent
  model overrides for build, plan, explore, and compaction agents
  (\texttt{agent/agent.ts:78--233}). \\
Plan/Act mode-based &
  \textsf{Cline} &
  Independent model per mode (\texttt{api/index.ts:76--149}).
  LLM switches modes via tool calls. \\
Per-attempt cycling &
  \textsf{SWE-agent}, \textsf{AutoCodeRover} &
  Different models for retry attempts.
  \textsf{SWE-agent}: round-robin through \texttt{agent\_configs}
  (\texttt{agents.py:303--319}).
  \textsf{AutoCodeRover}: round-robin through \texttt{model\_names}
  (\texttt{inference.py:98--114}). \\
Safety-focused (Guardian) &
  \textsf{Codex CLI} &
  Separate model (\texttt{gpt-5.4}) evaluates tool call risk
  (\texttt{guardian.rs}). \\
Task-based dual-model &
  \textsf{Prometheus} &
  Advanced model for reasoning (analysis, editing, patch selection);
  base model for mechanical tasks (retrieval, verification)
  (\texttt{llm\_service.py:23--38}). Hard-coded per graph node. \\
Actor-critic &
  \textsf{Moatless Tools} &
  Value function (potentially different model) scores nodes
  (\texttt{value\_function/base.py}). \\
Classifier chain &
  \textsf{Gemini CLI} &
  7-layer priority routing: fallback $\to$ override $\to$
  approval mode $\to$ Gemma classifier $\to$ LLM classifier
  $\to$ numerical classifier $\to$ default
  (\texttt{modelRouterService.ts:39--67}). Optional local
  Gemma model for client-side routing. \\
\bottomrule
\end{tabularx}
\end{table*}

\paragraph{Cost optimization is the primary driver.}
Across the agents that use multiple models, the dominant motivation is
cost: routing mechanical tasks to cheaper models while reserving
expensive ones for reasoning. \textsf{Aider}'s ``weak model'' handles
summarization and commit messages; in
architect mode, a third ``editor model'' receives the plan and generates
edits. \textsf{OpenCode} extends this role-based approach with
per-sub-agent model overrides.

\textsf{Prometheus} applies the same principle at a finer granularity.
Each node in its LangGraph state machine is hard-coded to use either
the ``advanced'' or ``base'' model.
Reasoning-heavy nodes (analysis,
editing, and patch selection, the last of which makes 10
majority-vote LLM calls) use the advanced
model, while mechanical nodes (knowledge graph traversal, test
execution) use the base model. Because the routing is per-node rather
than per-role, the cost savings scale with the proportion of
mechanical steps in the graph.

\paragraph{\textsf{Gemini CLI}'s classifier chain.}
\textsf{Gemini CLI}'s 7-layer routing strategy
is the most complex routing
mechanism observed. Each layer implements a different strategy, and the
first to produce a decision wins, so simple cases resolve cheaply while
ambiguous ones fall through to more sophisticated classifiers. The most
distinctive layer is the optional \texttt{GemmaClassifierStrategy},
which runs a lightweight Gemma model~\citep{gemma2024} locally for client-side routing
decisions, avoiding an API call just to select a model. No other agent
performs client-side model selection.

\paragraph{Actor-critic in tree search.}
\textsf{Moatless Tools}' value function
(\texttt{value\_function/base.py}) assigns numeric rewards to search
tree nodes. When configured to use a different model than the action
agent, the result is an actor-critic architecture: one model generates
actions while a separate model evaluates them. No other agent uses
separate models for generation and evaluation; \textsf{DARS-Agent}'s
LLM critic performs a similar role
but uses the same model for both, meaning the critic shares the
generator's biases.

\paragraph{Other routing strategies.}
Two agents use per-attempt model cycling: \textsf{SWE-agent} and
\textsf{AutoCodeRover} both rotate through different model
configurations on retry attempts,
betting that a model that failed on one
attempt may succeed if a different model tries the same task from
scratch. \textsf{Cline} routes by mode rather than by role: its plan
and act modes each use an independently configured model
and the LLM switches between modes
via tool calls. \textsf{Codex CLI} uses the most models of any agent in the corpus: the
primary model for code generation, a Guardian model (\texttt{gpt-5.4})
for safety evaluation of tool calls, and two
dedicated models for its memory extraction pipeline
(\texttt{gpt-5.1-codex-mini} for per-rollout extraction,
\texttt{gpt-5.3-codex} for consolidation;
Section~\ref{sec:results:memory}). The Guardian is the only instance in
the corpus of multi-model routing for safety rather than cost or
capability reasons.

\subsubsection{Persistent Memory}
\label{sec:results:memory}

Persistent memory (knowledge that survives across sessions) varies from
nonexistent to multi-tier extraction pipelines
(Table~\ref{tab:memory}). A clear division exists between agents
designed for benchmark evaluation and those designed for interactive
use. All five agents with no persistent
memory (\textsf{SWE-agent}, \textsf{OpenHands}, \textsf{AutoCodeRover}, \textsf{mini-swe-agent},
\textsf{DARS-Agent}) treat each task as independent, with no cross-task learning.
For benchmark-only agents (\textsf{SWE-agent}, \textsf{AutoCodeRover},
\textsf{mini-swe-agent}, \textsf{DARS-Agent}), this is expected; for
\textsf{OpenHands}, which is also widely used for interactive development
(53k stars, the second most in the corpus), the absence of persistent
memory is notable, though its microagent system loads static project
instructions that partially fill this role. \textsf{Agentless}
also lacks cross-task learning but supports pipeline resumability via cached
outputs and embedding indices, placing it between these two groups.
The five CLI agents
with persistent memory (\textsf{Aider}, \textsf{Cline}, \textsf{Gemini CLI}, \textsf{Codex CLI}, \textsf{OpenCode})
target interactive development where remembering project conventions
and past decisions has direct value. \textsf{Prometheus} is an outlier:
despite being a SWE-bench agent (Table~\ref{tab:agents}), it implements
multi-tier persistence (Neo4j, PostgreSQL, Athena) more characteristic
of an interactive tool, suggesting architectural ambitions beyond
benchmark evaluation. Among these, what ``memory'' means varies significantly:
\textsf{Cline} and \textsf{Gemini CLI} persist learned rules,
\textsf{Codex CLI} extracts and consolidates memories from past
sessions, while \textsf{OpenCode} persists full session state in SQLite,
including all messages, tool outputs, token usage, and costs
enabling interrupted sessions to be
resumed with complete context.

\begin{table*}[htbp]
\caption{Persistent memory strategies.}
\label{tab:memory}
\centering
\footnotesize
\begin{tabularx}{\textwidth}{p{2cm}p{2.3cm}X}
\toprule
Strategy & Agents & Mechanism \\
\midrule
None &
  \textsf{SWE-agent}, \textsf{OpenHands}, \textsf{AutoCodeRover},
  \textsf{mini-swe-agent}, \textsf{DARS-Agent} &
  No cross-session persistence. Each run starts fresh.
  Trajectory files are output artifacts, not consumed by future runs. \\
Pipeline resumability &
  \textsf{Agentless} &
  No cross-task learning, but JSONL outputs enable pipeline resumability
  (\texttt{--skip\_existing}), embedding indices persist via
  \texttt{--persist\_dir}, and pre-computed repo structures are cached via
  \texttt{PROJECT\_FILE\_LOC}. These are consumed by future runs of the
  same pipeline, not cross-task knowledge. \\
Config file loading &
  \textsf{Aider} (\texttt{.aider.conf.yml}) &
  Static, user-written configuration. Tags cache
  (\texttt{repomap.py:43, 217--265}) persists AST analysis
  across sessions as a performance optimization. \\
LLM-writable rules/memory &
  \textsf{Cline} (\texttt{.clinerules/}), \textsf{Gemini CLI} (\texttt{GEMINI.md}) &
  The LLM actively writes persistent instructions.
  \textsf{Cline}: \textsf{new\_rule} tool (\texttt{tools.ts:31}).
  \textsf{Gemini CLI}: \textsf{save\_memory} tool appends under
  ``Gemini Added Memories'' in \texttt{GEMINI.md}
  (\texttt{memoryTool.ts}). \\
Full session persistence &
  \textsf{OpenCode} &
  SQLite-backed session history with resume capability.
  All messages, tool outputs, token usage, and costs
  persist (\texttt{session/session.sql.ts:14--76}). \\
Background extraction pipeline &
  \textsf{Codex CLI} &
  Two-phase: Phase~1 extracts memories from recent rollouts
  (parallel, \texttt{gpt-5.1-codex-mini}); Phase~2 consolidates
  via sub-agent (\texttt{gpt-5.3-codex}). Usage-ranked, stale
  memories pruned (\texttt{memories/README.md}). \\
Multi-tier persistence &
  \textsf{Prometheus} &
  Athena (semantic memory service, HTTP API) + Neo4j
  (knowledge graph, 20 language ASTs) + PostgreSQL
  (LangGraph checkpoints). Memory-first retrieval with KG fallback
  (\texttt{context\_retrieval\_sub\-graph.py:159--163}). \\
\bottomrule
\end{tabularx}
\end{table*}

\paragraph{Static project instructions.}
Several agents in the ``None'' category do load static, user-written
project instructions that persist across sessions. \textsf{OpenHands}
reads \texttt{.openhands\_instructions} and \texttt{.cursorrules} from
the workspace via its microagent system, similar to \textsf{Aider}'s
\texttt{.aider.conf.yml}. These are not classified as persistent memory
because the agent never writes or updates them; they are static
configuration rather than learned knowledge.

\paragraph{LLM as memory author.}
\textsf{Cline} and \textsf{Gemini CLI} share a pattern where the LLM writes its own
persistent instructions. \textsf{Cline}'s \textsf{new\_rule} tool creates
\texttt{.clinerules} files; \textsf{Gemini CLI}'s
\textsf{save\_memory} tool appends to \texttt{GEMINI.md},
which additionally supports
\texttt{@path/to/file} references for composing instructions across
multiple files. These memories persist in the
project repository and are loaded into future sessions' system prompts. \textsf{Codex
CLI} takes a different approach: a background pipeline extracts and
consolidates memories without the LLM explicitly deciding what to
remember.

\paragraph{Cross-tool compatibility.}
\textsf{Cline} reads not only its own \texttt{.clinerules/} but also
\texttt{.cursorrules}, \texttt{.windsurfrules}, and \texttt{AGENTS.md}.
This pragmatic interoperability
reflects an emerging ecosystem where developers use multiple AI coding
tools with shared project-level instructions~\citep{ref12_configuring}.

\subsection{Cross-Cutting Themes}
\label{sec:results:cross-cutting}

Several findings span multiple taxonomy dimensions and do not reduce to
a single axis. These cross-cutting themes emerged from the open-ended
section of the analysis template (Section~\ref{sec:methodology:dimensions})
and represent architectural patterns or tradeoffs that manifest
differently across dimensions rather than constituting dimensions
themselves. They are presented separately because forcing them into the
dimensional framework would obscure their multi-dimensional nature.

\subsubsection{Sampling vs.\ Iteration}
\label{sec:results:sampling}

When an LLM-generated patch fails, there are two fundamentally
different ways to try again: generate another independent attempt
(sampling), or refine the failed attempt using feedback from the
failure (iteration). This distinction cuts across the control loop
taxonomy because it describes how agents handle the \emph{population}
of solution attempts, not the structure of any single attempt.

\textsf{Agentless} is the purest example of sampling: it generates
multiple patches independently and selects by majority voting
(Section~\ref{sec:results:loop-spectrum}). Each patch is generated from
the same context; no patch benefits from knowing that another failed.

Six of the nine LLM-driven agents (\textsf{OpenHands}, \textsf{Codex CLI},
\textsf{Gemini CLI}, \textsf{Cline}, \textsf{mini-swe-agent},
\textsf{OpenCode}) take the opposite approach: pure iteration. A
single attempt is refined through a feedback loop where each step sees
the results of previous steps. If a test fails, the agent sees the
error message and adjusts. This depth-first strategy bets that
feedback is more valuable than independence, and that a single guided
trajectory is more likely to converge on a correct solution than
multiple unguided ones.

Majority voting also appears in \textsf{Prometheus}, but at a
different level. Rather than generating patches independently,
\textsf{Prometheus} calls the advanced model 10 times on the same
prompt to \emph{select among} already-generated candidate patches
with early
stopping when the vote lead exceeds remaining votes. Here, voting
operates at the decision layer (choosing which patch to submit) rather
than the generation layer (creating patches).

\textsf{SWE-agent}'s \texttt{RetryAgent} occupies a middle ground
between sampling and iteration. It generates multiple complete
attempts, each a full depth-first trajectory with its own feedback
loop, and selects the best via a reviewer model.
Each attempt can use a different
model configuration. The result combines
iteration within each attempt (feedback-driven refinement) with
sampling across attempts (independent trajectories evaluated
post-hoc).

\subsubsection{Sub-Agent Delegation}
\label{sec:results:delegation}

Five agents support sub-agent spawning, where the primary agent
creates a secondary agent instance to handle a subtask. This matters
because it enables parallelism (working on multiple files
simultaneously) and specialization (delegating to agents with
different tool permissions or system prompts). However, these five
agents implement delegation through fundamentally different
mechanisms, revealing different assumptions about who should control
the delegation decision.

\textsf{Codex CLI} gives the LLM full control over delegation by
exposing it as a suite of tools: \textsf{spawn\_agent},
\textsf{send\_input}, \textsf{resume\_agent}, \textsf{wait}, and
\textsf{close\_agent}. The LLM decides
when to spawn a sub-agent, what task to assign it, and when to
collect results. Depth is limited by \texttt{agent\_max\_depth}, and
collaboration tools are disabled at maximum depth to prevent unbounded
recursion.

\textsf{OpenCode} takes a role-based approach: its \textsf{task} tool
spawns sub-agents with different specializations (build, plan,
explore, general), each with scaffold-enforced tool permissions.
The plan agent disables
file-editing tools (\texttt{edit}, \texttt{write}) for most paths but
retains bash access; the explore agent enables read-oriented tools plus
bash while denying write-oriented tools by default. The LLM chooses
which specialist to invoke, but the available specializations and their
capabilities are defined by the scaffold. This is a structural constraint, distinct from
\textsf{Cline}'s plan/act mode switching where the LLM itself decides
when to transition between modes via tool calls.

\textsf{OpenHands} integrates delegation into its event-sourced
architecture. A parent \texttt{AgentController} creates a child
controller via \texttt{AgentDelegateAction}
and forwards events to the
delegate. Because delegation flows through the same event stream as
all other actions, it is automatically captured in the agent's history
and subject to the same condensation and replay mechanisms.

\textsf{Cline} offers simpler tool-based delegation via
\textsf{new\_task} and
\textsf{use\_subagents}.
\textsf{Gemini CLI} supports sub-agents through its
\texttt{LocalAgentExecutor}, which runs a separate ReAct loop with its
own turn limits and deadline timer; uniquely, it includes a recovery
phase that gives the sub-agent one
final turn to produce output when the deadline expires.
\textsf{Prometheus}'s graph structure delegates implicitly through
subgraph nesting rather than explicit spawning.

\subsubsection{Online vs.\ Offline Selection}
\label{sec:results:online-offline}

Tree-search agents face two distinct selection problems: which branch
to explore next \emph{during} the search (online guidance), and which
completed solution to return \emph{after} the search finishes (offline
selection). These two problems can be solved by the same mechanism or
by different ones, with different tradeoffs.

\textsf{DARS-Agent} separates the two cleanly. During the search, an
LLM critic selects among candidate actions at each branch point;
this is online guidance that
shapes which parts of the search tree get explored. After the search
completes, a separately trained reviewer evaluates the finished
patches; this is offline selection that
chooses the final output. The separation means each component can be
optimized independently: the online critic needs to be fast (it runs
at every branch point), while the offline reviewer can be more
thorough (it runs once at the end).

\textsf{Moatless Tools} also separates the two components.
The value function assigns numeric
rewards during online MCTS simulation; a separate
discriminator evaluates
and selects the best completed trajectory offline. Unlike
\textsf{DARS-Agent}'s independently trained reviewer, both components
can be configured to use the same model, allowing the discriminator
to apply evaluation criteria consistent with the search guidance.

The contrast extends to how each agent extracts its final answer.
\textsf{DARS-Agent}'s leftmost-path extraction
always takes
\texttt{children[0]}, relying entirely on the online critic to have
directed the search toward good solutions. \textsf{Moatless Tools}'
discriminator actively re-evaluates all completed trajectories,
potentially selecting one that was not the most-visited during search.

\subsubsection{Ecosystem Maturity}
\label{sec:results:ecosystem}

The relationship between \textsf{DARS-Agent} and \textsf{SWE-agent}
illustrates the current state of ecosystem maturity. When
\textsf{DARS-Agent} needed to add tree-search capabilities on top of
\textsf{SWE-agent}'s ReAct loop, it copied and modified the entire
codebase rather than extending it: \texttt{agents.py} is a 700-line
copy of the original \textsf{SWE-agent} \texttt{Agent} class, and
\texttt{dars\_agent.py} is a parallel 1382-line reimplementation with
tree-search logic added. This fork-based reuse means bug fixes and
improvements in either project do not propagate to the other.

\textsf{mini-swe-agent} takes the opposite approach, reusing
\textsf{SWE-agent}'s execution environments (including its SWE-ReX
Docker and Modal backends) while defining its own
abstractions on top. It specifies agent, model, and environment as
Python Protocols, a form of
structural typing~\citep{pep544} where any object with the right methods
automatically conforms to the interface. This makes it possible to
swap in alternative implementations without modifying
\textsf{mini-swe-agent}'s code.

That fork-based and dependency-based reuse coexist for the same
upstream project suggests the ecosystem has not yet stabilized around
clean extension points. \textsf{Moatless Tools}' dual-flow architecture
(Section~\ref{sec:results:loop-spectrum}) offers a glimpse of what such
stabilization could look like: a clean separation between per-step logic
and orchestration strategy that makes switching between sequential and
tree-search modes a configuration choice.

\subsubsection{IDE as Architecture}
\label{sec:results:ide}

\textsf{Cline} is the only agent in the corpus that runs as an IDE
extension rather than a standalone CLI or server~\citep{cline}. This architectural
choice gives it access to a category of context that CLI agents cannot
obtain: the IDE's own understanding of the project.

For example, the \texttt{@problems} mention pulls from VS Code's
diagnostic API, giving the agent access to the same type errors,
linting warnings, and build failures that a developer sees in the
editor's ``Problems'' panel. The \texttt{@terminal} mention accesses
integrated terminal output, and commands execute in visible terminal
panels via the VS Code terminal
API, so the user can watch
commands run in real time. The
\texttt{FileContextTracker} monitors which files the model has
seen, read, or modified across turns, detecting external modifications
to prevent stale context during diff editing.

The tradeoff is platform lock-in. These capabilities depend on VS
Code's extension API, and \textsf{Cline}'s core features (diagnostics
integration, file watching, checkpoint git integration) cannot run
outside VS Code. The codebase shows signs of ongoing decoupling (a
standalone terminal manager, a CLI mode), but the richest context
sources remain VS Code-dependent.

\section{Discussion}
\label{sec:discussion}

The results presented in Section~\ref{sec:results} describe 12 dimensions
across three architectural layers, each exhibiting a range of strategies
observed in 13 open-source coding agents. This section interprets those
findings along three lines corresponding to the contributions stated in
Section~\ref{sec:intro}: what the spectral, compositional character of
the design space means for how researchers classify agents
(Sections~\ref{sec:discussion:spectra}--\ref{sec:discussion:convergence}),
what the taxonomy enables for researchers studying agent behavior and
practitioners building new scaffolds
(Sections~\ref{sec:discussion:interface}--\ref{sec:discussion:design}), and
what the evidence base reveals about ecosystem maturity and evaluation
methodology
(Sections~\ref{sec:discussion:evaluation}--\ref{sec:discussion:ecosystem}).

\subsection{Spectra, Not Categories}
\label{sec:discussion:spectra}

The most persistent finding across all 12 dimensions is that scaffold
architectures resist discrete classification. Prior taxonomies of LLM agents
organize systems into categories defined by abstract capabilities: tool-using,
memory-augmented, planning, reflective~\citep{masterman2024landscape,
nowaczyk2025agentic}. Every agent in the present corpus qualifies for every
one of these categories, yet their implementations differ in ways that these
labels cannot express. The source-code analysis reveals that the variation is
better captured as continuous spectra: control loops range from fixed
pipelines to full MCTS (Section~\ref{sec:results:loop-spectrum}), tool counts
range from 0 to 37 (Section~\ref{sec:results:toolset}), context compaction
ranges from none to LLM-initiated compression
(Section~\ref{sec:results:compaction}), and state management ranges from
destructive overwrite to event sourcing (Section~\ref{sec:results:state}).
Within each spectrum, agents occupy distinct positions that reflect genuine
architectural tradeoffs rather than arbitrary implementation choices.

This spectral character has a structural explanation: the loop primitives
identified in Section~\ref{sec:results:loop-spectrum} (ReAct, generate-test-repair,
plan-execute, multi-attempt retry, tree search) function as composable
building blocks that agents freely layer and nest. Because these primitives
compose freely, the space of possible architectures is combinatorial rather
than categorical. Assigning a single label (``ReAct agent,'' ``pipeline
agent'') to a system that layers multiple primitives obscures the design
decisions that actually differentiate it.

For researchers, this finding suggests that evaluating scaffold dimensions
independently may be more informative than classifying whole agents. A study
comparing ``ReAct agents'' against ``pipeline agents'' conflates loop
topology, loop driver, tool set design, and context management into a single
binary. Decomposing along the dimensions identified here would allow more
precise attribution of behavioral differences to specific architectural
choices. For practitioners, the composability of loop primitives suggests that
design decisions may be more orthogonal than they appear: \textsf{Moatless
Tools} demonstrates that tree search can be layered over an existing
per-step agent without rewriting the agent logic
(Section~\ref{sec:results:loop-spectrum}), though whether all dimension
combinations are equally feasible remains an open empirical question.

\subsection{Convergence and Divergence}
\label{sec:discussion:convergence}

Across the 12 dimensions, some show strong convergence among agents while
others show wide divergence, and the pattern is informative. The converging
dimensions tend to reflect constraints that are external to the scaffold
designer: tool capability categories converge on reading, searching, editing,
and executing code (Section~\ref{sec:results:toolset}) because these are the
operations that software engineering tasks require, regardless of
architectural philosophy. Edit format is trending toward string replacement
(Section~\ref{sec:results:edit-format}) because LLMs produce more reliable
edits with exact string matching than with line-number-based or unified-diff
formats whether through independent
discovery or imitation of successful designs.
Execution isolation converges on Docker containers for benchmark agents
(Section~\ref{sec:results:isolation}) because autonomous code execution
without sandboxing is unacceptable for unattended evaluation. These
convergences reflect solved problems or hard constraints: the design space has
been explored, and practitioners have settled on solutions that work.

The diverging dimensions tell a different story. Context compaction
(Section~\ref{sec:results:compaction}) exhibits seven distinct strategies
across 13 agents, from no management at all to LLM-initiated compression with
verification probes. State management (Section~\ref{sec:results:state})
ranges from destructive overwrite to event sourcing, with tree-structured,
graph-scoped, and database-backed variants between them. Multi-model routing
(Section~\ref{sec:results:routing}) spans single-model simplicity to
seven-layer classifier chains. These dimensions diverge because they address
open design questions where no dominant solution has emerged. Context
compaction, for instance, requires balancing information preservation against
token cost, with the optimal tradeoff depending on task length, model
capability, and cost tolerance in ways that no single strategy resolves.
State management involves similar tradeoffs between simplicity, auditability,
and support for branching exploration. The divergence on these dimensions is
not noise; it reflects genuine uncertainty about the best approach.

This pattern has practical implications. The converging dimensions are
candidates for standardization: a shared tool protocol covering the four
capability categories (as the Model Context Protocol begins to attempt) would
reduce duplicated effort without constraining architectural innovation.
The diverging dimensions, conversely, are where research investment is most
needed. The diversity of context compaction strategies, in particular,
suggests that no existing approach fully solves the ``token
snowball'' problem identified by \citet{fan2025sweeffi}, where
growing context from tool outputs degrades both performance and cost. The
range of observed strategies (prevention through structural bounding,
cure through summarization, and hybrid approaches) represents an active
design frontier.

\subsection{The Scaffold-Model Interface}
\label{sec:discussion:interface}

The taxonomy reveals that scaffold design mediates model capability in ways
that prior work has not systematically examined. The same underlying language
model behaves differently depending on how many tools it is presented (0 in
\textsf{Aider} versus 35 in \textsf{SWE-agent},
Section~\ref{sec:results:toolset}), how context is managed (full unfiltered
history in \textsf{mini-swe-agent} versus event-sourced views with
condensation in \textsf{OpenHands}, Section~\ref{sec:results:compaction}),
what loop structure surrounds it (single-pass pipeline in \textsf{Agentless}
versus feedback-driven iteration in \textsf{SWE-agent},
Section~\ref{sec:results:loop-spectrum}), and who drives the loop (user in
\textsf{Aider}, scaffold in \textsf{AutoCodeRover}, LLM in
\textsf{OpenHands}, Section~\ref{sec:results:loop-driver}). Each of these
scaffold-level decisions shapes what the model sees, what actions it can take,
and how its errors propagate or get corrected.

This observation has direct consequences for empirical studies of coding
agents. Trajectory analyses that compare agents using different
models~\citep{majgaonkar2025trajectories} cannot isolate whether an observed
behavioral difference stems from the scaffold or the model. The present
taxonomy provides the vocabulary to describe exactly which scaffold
dimensions differ between two agents, making it possible to design
controlled comparisons. For example, a study could compare agents with
identical tool sets but different loop strategies
(Section~\ref{sec:results:loop-spectrum}), or identical loops but different
compaction strategies (Section~\ref{sec:results:compaction}), holding the
model constant. \citet{ref4_archaware} called for
architecture-aware evaluation metrics that link internal components to
observable outcomes; the 12 dimensions identified here provide the
architectural variables that such metrics would need to control for.

The scaffold-model interface also explains why the loop driver dimension
(Section~\ref{sec:results:loop-driver}) is arguably the most fundamental
architectural distinction. As Section~\ref{sec:results:loop-driver}
documents, user-driven agents sidestep the localization bottleneck entirely,
while LLM-driven agents must solve it, making retrieval strategy a critical
co-design choice. The same model that performs well with user-curated context
may struggle when forced to navigate a repository autonomously. This
interaction illustrates why scaffold dimensions cannot be evaluated in
isolation; they form an interdependent design space.

\subsection{Implications for Agent Design}
\label{sec:discussion:design}

Several practical design lessons emerge from the taxonomy, though they should
be understood as patterns observed across 13 agents rather than prescriptive
recommendations.

\paragraph{Loop composition as a design strategy.}
Eleven of the 13 agents layer multiple loop primitives rather than relying
on a single control structure (Section~\ref{sec:results:loop-spectrum}).
Pure single-loop agents are rare: \textsf{Agentless} (pipeline only) and
\textsf{mini-swe-agent} (ReAct only) are the closest examples, and both are
deliberately minimalist. This pattern suggests that the ReAct loop, while
foundational, is typically insufficient on its own; layering a retry,
test-repair, or planning primitive on top addresses failure modes that a
single feedback loop cannot handle~\citep{shinn2023reflexion}.

\paragraph{Tool count and the capability-confusion tradeoff.}
Tool counts range from 0 to 37, yet the underlying capability categories
converge on four (read, search, edit, execute). This convergence at the
capability level despite divergence at the tool level suggests that the four
categories define a minimum viable toolset for autonomous coding agents.
Beyond this baseline, the tradeoff is between expressiveness and LLM
confusion: more specialized tools reduce the LLM's per-tool reasoning burden
but increase the action space it must navigate~\citep{yang2024sweagent}.
\textsf{Prometheus}'s per-node tool scoping
(Section~\ref{sec:results:toolset}) and \textsf{AutoCodeRover}'s phase
separation offer two strategies for managing this tradeoff, constraining the
tools visible at each decision point rather than presenting the full set at
every step.

\paragraph{Context compaction as an architectural requirement.}
Every agent that gives the LLM sustained autonomy must address context
growth. Section~\ref{sec:results:compaction} documents two philosophies:
prevention (bounding context structurally) and cure (compressing on demand).
Prevention avoids information loss but requires anticipating growth patterns;
cure is more flexible but risks lossy compression. The only agent with no
compaction strategy (\textsf{mini-swe-agent}) crashes when the context window
is exceeded, confirming that compaction is not optional for agents operating
beyond trivial task lengths.

\paragraph{Sub-agent delegation as an emerging capability.}
Five of the 13 agents support explicit sub-agent spawning through five distinct
mechanisms (Section~\ref{sec:results:delegation}); a sixth,
\textsf{Prometheus}, achieves implicit delegation through subgraph nesting
rather than explicit spawning. The diversity and the
absence of a dominant pattern suggest that delegation is an active design
frontier. The variation in who controls the delegation decision mirrors the
loop driver spectrum (Section~\ref{sec:results:loop-driver}), suggesting that
the autonomy-versus-control tradeoff recurs at every level of agent
architecture.

\subsection{Implications for Agent Evaluation}
\label{sec:discussion:evaluation}

As noted in Section~\ref{sec:methodology:scope}, SWE-bench comparisons
between agents confound scaffold design, model choice, and configuration in
a single metric. The taxonomy makes this confounding concrete.
As Section~\ref{sec:results:routing}
documents, agents use different models and different numbers of models;
\textsf{Codex CLI} routes to four distinct models while
\textsf{mini-swe-agent} uses one. Per-attempt model cycling in
\textsf{SWE-agent} and \textsf{AutoCodeRover}
(Section~\ref{sec:results:routing}) means that a single benchmark run may
involve multiple models, further complicating attribution. Controlling for
these differences requires the kind of architectural decomposition that the
present taxonomy provides but that existing evaluations do not perform.

The sampling-versus-iteration distinction
(Section~\ref{sec:results:sampling}) poses a particularly acute evaluation
challenge. \textsf{Agentless}'s independent sampling strategy and
\textsf{SWE-agent}'s iterative retry strategy represent fundamentally
different approaches to the same problem, but benchmark scores conflate
single-attempt quality with multi-attempt strategy. An agent that produces
mediocre individual patches but samples 40 of them and selects the best may
outperform an agent that produces strong individual patches through iterative
refinement. Separating these two capabilities in evaluation would require
reporting both single-attempt and multi-attempt metrics, a distinction the
taxonomy makes legible but current benchmarks do not enforce.

Architecture-aware evaluation need not require entirely new benchmarks. The
12 dimensions identified here suggest concrete controls that could be applied
within existing evaluation frameworks. For instance, fixing the tool set
(Section~\ref{sec:results:toolset}) while varying the control loop
(Section~\ref{sec:results:loop-spectrum}) would isolate the effect of loop
strategy on task success. Fixing the model while varying the scaffold would
isolate scaffold effects from model effects. The taxonomy provides the
variables; the evaluation methodology is a matter of experimental design.

\subsection{Ecosystem Maturity and Standardization}
\label{sec:discussion:ecosystem}

The state of reuse and modularity across the corpus reflects an ecosystem
that is innovating rapidly but has not yet stabilized around shared
abstractions. As Section~\ref{sec:results:ecosystem} documents, fork-based
and dependency-based reuse coexist for the same upstream project, indicating
that clean extension points have not yet emerged.

The convergence on tool capability categories without convergence on tool
interfaces (Section~\ref{sec:results:toolset}) suggests a specific
standardization opportunity. All LLM-driven agents need tools for reading,
searching, editing, and executing code, but each agent defines its own tool
schemas, parameter names, and output formats. A shared tool interface
protocol could reduce the integration cost of new tools without constraining
how scaffolds orchestrate them. The Model Context Protocol (MCP)~\citep{mcp2024}, already
supported by five agents in the corpus (\textsf{OpenHands},
\textsf{Codex CLI}, \textsf{Gemini CLI}, \textsf{Cline},
\textsf{OpenCode}), represents an early attempt at this kind of
standardization, though it operates at the transport layer rather than
defining semantic contracts for specific tool categories.

\textsf{Moatless Tools}' dual-flow architecture
(Section~\ref{sec:results:loop-spectrum}) offers a more focused model of what
modular scaffold design could look like. Its clean separation between the
per-step executor (\texttt{ActionAgent}) and the orchestration strategy
(\texttt{AgenticLoop} or \texttt{SearchTree}) means that adding a new
exploration strategy requires implementing a new orchestrator, not modifying
the agent logic. This separation of concerns is rare in the corpus; most
agents tightly couple their per-step logic with their orchestration strategy,
making it difficult to experiment with alternative control structures without
substantial refactoring. As the ecosystem matures, this kind of modular
separation may prove more valuable than tool protocol standardization,
because it addresses the architectural level where the most design variation
exists (Section~\ref{sec:discussion:convergence}).

\textsf{Cline}'s IDE integration (Section~\ref{sec:results:ide}) illustrates
a different facet of ecosystem evolution: the tradeoff between platform
coupling and capability richness. The IDE-native context that \textsf{Cline}
accesses (diagnostics, file-change tracking, terminal integration) is
unavailable to CLI agents, but comes at the cost of VS Code platform
lock-in. For the ecosystem as a whole, this tension may resolve through
protocol-level abstraction (providing IDE-quality context via standardized
APIs) rather than platform convergence, but no such abstraction exists today.

\section{Threats to Validity}
\label{sec:threats}

This section organizes threats to validity following the standard framework for
empirical software engineering studies~\citep{runeson2009guidelines}: construct
validity (whether the study measures what it claims to measure), internal
validity (whether the findings follow from the data), external validity
(whether the findings generalize beyond this study), and reliability (whether
the study can be reproduced).

\subsection{Construct Validity}

The primary construct validity threat is single-author bias. All 13
agent analyses were conducted by a single author (with LLM-assisted code
navigation, as described in Section~\ref{sec:methodology:procedure}),
meaning that dimension classifications, evidence selection, and
cross-agent comparisons reflect one person's interpretation of the source
code. Two mitigations
partially address this threat. First, every taxonomic claim in
Section~\ref{sec:results} is grounded in specific file paths and line
numbers pinned to commit hashes (Appendix~\ref{sec:appendix:commits}),
making each claim independently verifiable against the source code. A
post-hoc verification pass checked 296 extracted claims against the
cloned repositories, confirming 267, correcting 19 (primarily line number
offsets from code evolution between analysis and verification dates), and
accepting 10 as minor simplifications (e.g., describing a multi-step
process in fewer steps than the code implements, or attributing a
behavior to a single function when it spans two). The verification pass
was performed by the same researcher who conducted the original analysis;
while self-verification is weaker than independent review, the
commit-pinned evidence trail makes independent verification
straightforward.
Second, the analysis template
separates observation (what the code does), classification (how it maps
to a dimension), and evidence (file path and line number), following
case study reporting guidelines~\citep{runeson2009guidelines}. This
separation makes it possible for a reader to evaluate the classification
independently of the observation. Nonetheless, the study would benefit
from independent replication by a second analyst, particularly for
dimensions where judgment calls are required (for example, whether
\textsf{Prometheus}'s graph-scoped state management constitutes a form
of structural compaction or a separate architectural pattern).

A second construct validity threat concerns the dimension framework
itself. Although the nine analysis dimensions were derived iteratively
through open coding~\citep{strauss1998basics} during a pilot analysis of
two architecturally contrasting agents
(Section~\ref{sec:methodology:dimensions}), the pilot agents
(\textsf{Aider} and \textsf{OpenHands}) may not have surfaced all
relevant dimensions. The open-ended tenth section of the analysis
template was designed to capture observations outside the predefined
dimensions, and it produced 47 cross-cutting findings that informed the
final taxonomy (Section~\ref{sec:results:cross-cutting}). However,
dimensions that neither pilot agent exhibits and that no subsequent
agent made salient could still be missing. For example, the taxonomy
does not include a dimension for prompt engineering strategy (how system
prompts are constructed and varied). While prompt templates are visible
in source code, the dimension was excluded for scope: analyzing prompt
structure, length, few-shot examples, and persona instructions across
13 agents would constitute a study in its own right, and the
architectural impact of prompt differences (as opposed to scaffold
differences) cannot be assessed without runtime experimentation. This
exclusion is deliberate but means that an important aspect of scaffold
design is not captured.

\subsection{Internal Validity}

Internal validity concerns whether the observed patterns genuinely
reflect architectural relationships rather than confounds. Two threats
are relevant.

First, the taxonomy describes source code at pinned commits, but several
agents were under active development during the analysis period
(Section~\ref{sec:methodology:procedure}). Architectural features may
have been added, removed, or significantly refactored between the
analyzed commit and the current version. Pinning to specific commits
ensures reproducibility of the reported findings but means the taxonomy
is a snapshot, not a live description. The commit hashes are listed in
Appendix~\ref{sec:appendix:commits} so that readers can assess how much
each agent has evolved since analysis.

Second, some dimensions may not be as independent as the taxonomy
implies. Section~\ref{sec:discussion:interface} notes that loop driver
and retrieval strategy are correlated: scaffold-driven agents tend to
invest in retrieval infrastructure while LLM-driven agents rely on
general-purpose tools. Similar correlations may exist between other
dimensions (for example, between tool discovery strategy and tool count,
or between state management and context compaction). The taxonomy
presents dimensions as independent axes, but in practice they form an
interdependent design space where choices on one dimension constrain
options on others. The cross-cutting themes in
Section~\ref{sec:results:cross-cutting} capture some of these
interdependencies, but a full analysis of dimension interactions is
beyond the scope of this study.

\subsection{External Validity}

Three threats limit the generalizability of the findings.

First, the corpus is restricted to open-source agents with readable
source code (Section~\ref{sec:methodology:selection}). Proprietary
coding agents (GitHub Copilot Workspace, Cursor's AI backend, Windsurf)
and agents with compiled or obfuscated source code (Claude Code) are
excluded because their scaffolding is not publicly inspectable. This
introduces a survivorship bias: open-source agents may systematically
differ from proprietary agents in design choices driven by business
constraints, proprietary model access, or different optimization targets
(user experience versus benchmark scores). The taxonomy should therefore
be understood as describing the open-source design space, not the full
design space of coding agents.

Second, the corpus of 13 agents, while covering a range of architectural
strategies, is not exhaustive. Agents released after the analysis period
or agents that did not meet the inclusion criteria may exhibit
architectural patterns not represented in the taxonomy. The study aims
for analytical generalizability~\citep{yin2018case} (the dimensions and spectra should be
useful for characterizing new agents) rather than statistical
generalizability (the distribution of agents across dimension positions
is not claimed to be representative of any population). As the ecosystem
evolves, both new dimensions and new positions on existing dimensions
are likely to emerge.

Third, the corpus is dominated by agents targeting Python-language
repositories, primarily because SWE-bench (the dominant evaluation
benchmark) uses Python projects exclusively. Agents designed for
multi-language or non-Python ecosystems may face different architectural
constraints (for example, different AST parsing requirements, different
build and test toolchains, or different dependency resolution patterns)
that could produce architectural variation not observed here~\citep{jimenez2024swebench,xu2025swecompass}.
\textsf{Prometheus}'s 20-language tree-sitter support and
\textsf{SWE-agent}'s language-agnostic shell-based approach suggest that
some agents already address this limitation, but the analysis does not
systematically evaluate how architectural choices vary across target
languages.

\subsection{Reliability}

The primary reliability threat is reproducibility. The fully specified
template (Section~\ref{sec:methodology:procedure}) and pinned commit
hashes (Appendix~\ref{sec:appendix:commits}) enable a second analyst to
arrive at substantially similar observations, though classification
judgments (where to place an agent on a continuous spectrum) may differ.
The complete analysis documents enable independent scrutiny.

A secondary concern is that static source-code analysis misses runtime
behavior. Some architectural features (MCP tool discovery in practice,
whether configurable features like \textsf{Moatless Tools}' pluggable
selector are used in typical deployments) may only be visible at
runtime. The taxonomy describes architectural capability, not observed
runtime behavior.

The corpus is also heavily concentrated in Python-implemented agents
(10 of 13); the three TypeScript agents (\textsf{Cline},
\textsf{Gemini CLI}, \textsf{OpenCode}) may use language idioms
(event-driven architectures, module systems) that a Python-oriented
analysis framework is less attuned to. The analysis template was
designed to be language-agnostic, but subtle biases in what the analyst
notices cannot be ruled out.

Finally, the CLI/SWE-bench category distinction used throughout the
paper (Table~\ref{tab:agents}) is a simplification. Some agents
straddle categories: \textsf{OpenHands} is categorized as SWE-bench
but is also widely used as an interactive development tool, and several
SWE-bench agents can be run interactively. The distinction is used
descriptively (to contextualize design choices) rather than analytically
(as a dimension of the taxonomy), but readers should not interpret it as
a rigid boundary.

\section{Conclusion}
\label{sec:conclusion}

This paper presented a source-code-level architectural taxonomy of 13
open-source coding agent scaffolds, organized into three layers (control
architecture, tool and environment interface, resource management) and 12
dimensions. Every taxonomic claim is grounded in file paths and line numbers
from cloned repositories at pinned commits, providing an evidence base that
can be independently verified and extended as the ecosystem evolves.

Three findings emerge from the analysis. First, scaffold architectures are
better characterized as positions along continuous spectra than as instances
of discrete types. Control strategies range from fixed pipelines to full
Monte Carlo Tree Search; tool counts range from 0 to 37; context compaction
spans seven distinct strategies; state management ranges from destructive
overwrite to event sourcing. Prior capability-based taxonomies that classify
agents as ``tool-using'' or ``planning'' cannot distinguish between systems
that differ fundamentally on these dimensions. Second, the loop primitives
that underlie control architectures (ReAct, generate-test-repair,
plan-execute, multi-attempt retry, tree search) function as composable
building blocks: 11 of the 13 agents layer multiple primitives rather than
relying on a single control structure. This compositionality means the design
space is combinatorial rather than categorical, and assigning a single
architectural label to a system obscures the design decisions that
differentiate it. Third, the dimensions themselves exhibit a pattern of
convergence on externally constrained choices (tool capability categories,
edit formats, execution isolation) and divergence on open design questions
(context compaction, state management, multi-model routing), suggesting where
the design space has stabilized and where research investment is most needed.

Several directions for future work follow from the taxonomy.
The most direct extension is controlled experimentation: the 12 dimensions
identify specific architectural variables that can be isolated while holding
others constant. For example, comparing agents with identical tool sets but
different loop strategies, or identical loops but different compaction
strategies, with the model held constant, would enable causal attribution of
performance differences to scaffold design rather than to the model or
configuration confounds that current benchmark comparisons cannot
disentangle (Section~\ref{sec:discussion:evaluation}). The taxonomy provides
the variables; designing the experiments is the next step.

A second direction is longitudinal analysis. The taxonomy describes a
snapshot of 13 agents at pinned commits. Repeating the analysis at later
commits would reveal how scaffold architectures evolve: whether converging
dimensions continue to converge, whether diverging dimensions stabilize, and
whether new dimensions emerge as the ecosystem matures. The commit-pinned
methodology makes such longitudinal comparison straightforward.

Third, extending the corpus to proprietary agents (when architectural
details become available through documentation or reverse engineering) and to
agents targeting languages beyond Python would test the generalizability of
the observed spectra (Section~\ref{sec:threats}). The dimension framework is
designed to be language- and platform-agnostic, but whether the specific
positions observed here generalize to different ecosystems remains an open
question.

Finally, the taxonomy enables the architecture-aware evaluation metrics that
prior work has called for~\citep{ref4_archaware} but could not implement
without architectural documentation. Linking specific dimension positions
(loop strategy, compaction approach, tool set design) to observable outcomes
(task success, token cost, trajectory length) would move the field from
system-level leaderboards toward component-level understanding of what makes
coding agents effective.

\bibliographystyle{plainnat}
\bibliography{references}

\appendix

\section{Candidate Agent Corpus}
\label{sec:appendix:candidates}

Table~\ref{tab:candidates} lists the full pool of 22 candidate agents
considered for this study, along with the inclusion criterion each excluded
agent failed. The three inclusion criteria are defined in
Section~\ref{sec:methodology:selection}.

\begin{table*}[htbp]
\caption{Full candidate pool. Agents are grouped by disposition: the 13
included agents appear first (ordered as in Table~\ref{tab:agents}),
followed by the 9 excluded agents grouped by exclusion criterion.}
\label{tab:candidates}
\centering\footnotesize
\begin{tabularx}{\textwidth}{llX}
\toprule
\textbf{Agent} & \textbf{Disposition} & \textbf{Exclusion rationale} \\
\midrule
\textsf{Gemini CLI}         & Included & --- \\
\textsf{OpenHands}          & Included & --- \\
\textsf{Aider}              & Included & --- \\
\textsf{Cline}              & Included & --- \\
\textsf{SWE-agent}          & Included & --- \\
\textsf{Codex CLI}          & Included & --- \\
\textsf{OpenCode}           & Included & --- \\
\textsf{Agentless}          & Included & --- \\
\textsf{AutoCodeRover}      & Included & --- \\
\textsf{Moatless Tools}     & Included & --- \\
\textsf{Prometheus}         & Included & --- \\
\textsf{DARS-Agent}         & Included & --- \\
\textsf{mini-swe-agent}     & Included & --- \\
\midrule
\multicolumn{3}{l}{\emph{Excluded: not coding-specific (Criterion~1)}} \\
\textsf{Open Interpreter}   & Excluded &
  General-purpose code execution framework. No codebase navigation,
  patch application, or git integration. \\
\textsf{Deep Agents}        & Excluded &
  General-purpose LangGraph agent harness. No git integration, no
  diff/patch application, no test runner, no AST tooling. \\
\textsf{MetaGPT}            & Excluded &
  Multi-agent orchestration framework; unit of analysis is agent
  coordination, not individual scaffold architecture. \\
\textsf{CrewAI}             & Excluded &
  General-purpose multi-agent orchestration platform; same rationale
  as \textsf{MetaGPT}. \\
\midrule
\multicolumn{3}{l}{\emph{Excluded: no readable source code (Criterion~2)}} \\
\textsf{Claude Code}        & Excluded &
  Distributed as a compiled npm binary; no published source
  repository. \\
\textsf{MASAI}              & Excluded &
  Repository contains only a README linking to the paper; no
  implementation code released. \\
\textsf{Copilot Workspace}  & Excluded &
  Proprietary; scaffold code not publicly inspectable. \\
\textsf{Cursor}             & Excluded &
  Commercial AI code editor; scaffold code not publicly inspectable. \\
\textsf{Windsurf}           & Excluded &
  Commercial AI code editor; scaffold code not publicly inspectable. \\
\bottomrule
\end{tabularx}
\end{table*}

\section{Pinned Commit Hashes}
\label{sec:appendix:commits}

Table~\ref{tab:commits} lists the repository URL and pinned commit hash for
each analyzed agent. All file paths and line numbers cited in
Section~\ref{sec:results} refer to these specific commits. Readers can
clone each repository and check out the listed commit to reproduce or
verify any claim in this paper.

\begin{table*}[htbp]
\caption{Pinned commit hashes for all 13 analyzed agents. Commits are
listed in the same order as Table~\ref{tab:agents}.}
\label{tab:commits}
\centering\footnotesize
\begin{tabularx}{\textwidth}{llX}
\toprule
\textbf{Agent} & \textbf{Commit hash} & \textbf{Repository URL} \\
\midrule
\textsf{Gemini CLI}
  & \texttt{dd8d4c98b3} & \url{https://github.com/google-gemini/gemini-cli} \\
\textsf{OpenHands}
  & \texttt{922e3a2431} & \url{https://github.com/OpenHands/OpenHands} \\
\textsf{Aider}
  & \texttt{861a1e4d15} & \url{https://github.com/Aider-AI/aider} \\
\textsf{Cline}
  & \texttt{71e312e92a} & \url{https://github.com/cline/cline} \\
\textsf{SWE-agent}
  & \texttt{e72a7e4660} & \url{https://github.com/SWE-agent/SWE-agent} \\
\textsf{Codex CLI}
  & \texttt{9dba7337f2} & \url{https://github.com/openai/codex} \\
\textsf{OpenCode}
  & \texttt{f54abe58cf} & \url{https://github.com/anomalyco/opencode} \\
\textsf{Agentless}
  & \texttt{5ce5888b9f} & \url{https://github.com/OpenAutoCoder/Agentless} \\
\textsf{AutoCodeRover}
  & \texttt{585d3e639a} & \url{https://github.com/AutoCodeRoverSG/auto-code-rover} \\
\textsf{Moatless Tools}
  & \texttt{011ead57a5} & \url{https://github.com/aorwall/moatless-tools} \\
\textsf{Prometheus}
  & \texttt{b1c722be02} & \url{https://github.com/EuniAI/Prometheus} \\
\textsf{DARS-Agent}
  & \texttt{eab35168a9} & \url{https://github.com/vaibhavagg303/DARS-Agent} \\
\textsf{mini-swe-agent}
  & \texttt{6f1b196616} & \url{https://github.com/SWE-agent/mini-swe-agent} \\
\bottomrule
\end{tabularx}
\end{table*}

\end{document}